\documentclass[sn-mathphys]{sn-jnl_edited}

\jyear{2023}%

\theoremstyle{thmstyleone}%
\theoremstyle{thmstyletwo}%

\theoremstyle{thmstylethree}%

\newcommand{\covid}{\textsc{covid}-19}
\newcommand{\sars}{\textsc{sars}-\textsc{c}o\textsc{v}-2}

\usepackage{bm}

\usepackage{siunitx} 

\usepackage{pdflscape}
\usepackage{gensymb}

\usepackage{listings}

\definecolor{codegreen}{rgb}{0,0.6,0}
\definecolor{codegray}{rgb}{0.5,0.5,0.5}
\definecolor{codepurple}{rgb}{0.58,0,0.82}
\definecolor{backcolour}{rgb}{0.95,0.95,0.92}

\lstdefinestyle{mystyle}{
    backgroundcolor=\color{backcolour},   
    commentstyle=\color{codegreen},
    keywordstyle=\color{magenta},
    numberstyle=\tiny\color{codegray},
    stringstyle=\color{codepurple},
    basicstyle=\ttfamily\footnotesize,
    breakatwhitespace=false,         
    breaklines=true,                 
    captionpos=b,                    
    keepspaces=true,                 
    numbers=left,                    
    numbersep=5pt,                  
    showspaces=false,                
    showstringspaces=false,
    showtabs=false,                  
    tabsize=2
}

\usepackage{lineno}
\usepackage{mhchem}

\graphicspath{{fig/}}

\begin{document}

\title{pySODM: Simulating and Optimizing Dynamical Models in Python 3}


\author*[1,2]{\fnm{Tijs W.}
\sur{Alleman}}\email{tijs.alleman@ugent.be}

\author[3]{\fnm{Christian}
\sur{Stevens}}

\author[1]{\fnm{Jan M.}
\sur{Baetens}}

\affil[1]{\orgdiv{KERMIT}, \orgname{Department of Data Analysis and Mathematical Modelling, Ghent University}, \orgaddress{\street{Coupure Links 653}, \city{Ghent}, \postcode{9000}, \country{Belgium}}}

\affil[2]{\orgdiv{BIOMATH}, \orgname{Department of Data Analysis and Mathematical Modelling, Ghent University}, \orgaddress{\street{Coupure Links 653}, \city{Ghent}, \postcode{9000}, \country{Belgium}}}

\affil[3]{\orgdiv{SynBioC}, \orgname{Department of Green Chemistry and Technology, Ghent University}, \orgaddress{\street{Coupure Links 653}, \city{Ghent}, \postcode{9000}, \country{Belgium}}}


\maketitle

\noindent\textbf{Abstract} In this work we present our generic framework to construct, simulate and calibrate dynamical systems in Python 3. Its goal is to reduce the time it takes to implement a dynamical system with $n$-dimensional states represented by coupled ordinary differential equations (ODEs), simulate the system deterministically or stochastically, and, calibrate the system using $n$-dimensional data. We demonstrate our code's capabilities by building three models in the context of two case studies. First, we forecast the yields of the enzymatic esterification reaction of D-glucose and lauric acid, performed in a continuous-flow, packed-bed reactor. The model yields a satisfactory description of the reaction yields under different flow rates and can be applied to design a viable process. Second, we build a stochastic, age-stratified model to make forecasts on the evolution of influenza in Belgium during the 2017--2018 season. Using only limited data, our simple model was able to make a fairly accurate assessment of the future course of the epidemic. By presenting real-world case studies from two scientific disciplines, we demonstrate our code's applicability across domains.\\


\noindent\textbf{Word count} 4900 words (main text) excluding captions of figures and tables.\\

\noindent\textbf{Availability of Data and Code} The source code of pySODM is freely available on GitHub: \url{https://github.com/twallema/pySODM}. A documentation website is available on on \url{https://twallema.github.io/pySODM}. All data necessary to reproduce the case studies shown in this work are available on GitHub.


\section{Introduction}\label{sec:introduction}

Differential equations are used to describe a wide variety of processes and are the workhorses of most applied mathematics, physics, and engineering \citep{Goriely2018, Smith2016, Villaverde2021}. Both from personal experience, as well as described by Villaverde et al., 2021 \citep{Villaverde2021}, a typical simulation \& calibration workflow constitutes the following steps: 1) Translate a real-world phenomenon into a set of differential equations. Analyze their structural identifiability (if possible), and implement them using a programming language. 2) Use a set of experimental data to calibrate some of the model's parameters. 3) Verify the goodness of fit. 4) Analyze the distributions of the calibrated parameters to asses their practical identifiability. 5) Use the model to gain additional insights into the process or make projections beyond the calibrated range.\\

\noindent The goal of pySODM is to reduce the time needed to go through the aforementioned workflow. It facilitates the implementation of a dynamical system with $n$-dimensional states represented by coupled ordinary or partial differential equations (ODEs), the deterministic or stochastic simulation of the system, the variation of model parameters during a simulation, and, the calibration to $n$-dimensional data. An overview of pySODM's features is provided in Table \ref{tab:overview_features}.\\

\noindent Established low-level interfaces to integrate sets of ODEs (\texttt{scipy.integrate}, \citep{Scipy2020}), simulate stochastic jump processes, known as stochastic simulation algorithms (SSAs), Doob's method, Gillespie methods, or Kinetic Monte Carlo methods across different fields of science \citep{Gillespie1977, Gillespie2001}, perform frequentist optimizations of model parameters using Particle Swarm Optimization \citep{Kennedy1995} or the Nelder-Mead Simplex algorithm \citep{Nelder1965}, and, perform Bayesian inference of model parameters (\texttt{emcee.EnsembleSampler}, \citep{emcee2013}), are readily available in Python 3. pySODM overcomes two problems preventing an efficient workflow. First, convenient simulation features, such as time-dependent model parameters, are missing in the aforementioned implementations. Second, to integrate these third-party implementations in the aforementioned workflow, an easy-to-use, uniform way of storing and indexing simulation results is needed. To this end, pySODM formats simulation results using the \texttt{xarray.Dataset} \citep{hoyer2017}, which also eases scripting pySODM models with third-party software and applications, such as SAlib \citep{Herman2017} for sensitivity analysis. A conceptual representation of pySODM is shown in Figure \ref{fig:conceptualization}. \\

\noindent In Python 3, the closest alternative to pySODM is pyGOM \citep{pyGOM2018}. It can be used to solve systems of ODEs deterministically or stochastically and can be used to construct an objective function for optimization. The key difference between both packages is that pyGOM has users define their system using symbolic transitions whereas pySODM has users define a function to compute the model's differentials. The use of symbolic transitions is a more high-level approach and offers three advantages. First, the use of symbolic transitions is more adept for novice users. Second, properties of the system, such as bifurcation and structural identifiability can readily be analyzed. Second, gradient information is available for optimization algorithms resulting in a more efficient search strategy. However, from our experience, the use of symbolic transitions imposes a limit on the attainable model complexity. As an example, it is not possible to transfer individuals to a vaccinated state in a disease transmission model based on real-world incidence data using symbolic transitions. pySODM was designed from the start to offer users maximum flexibility and facilitate the construction of arbitrarily complex models. Another advantage of pySODM is the ability to dynamically vary model parameters (time-dependent model parameters) while pyGOM, to our knowledge, does not. In the context of \sars{} dynamic transmission modeling, these time-dependent model parameters were used to inform the number of vaccinated individuals that needed to be transferred to a vaccinated state during the \covid{} pandemic \citep{Alleman2023a}.\\

\noindent The most comprehensive alternative to pySODM is the SciML ecosystem in Julia \citep{SciML}, which bundles several packages to support a similar modeling and simulation workflow. The \texttt{DifferentialEquations.jl} suite offers tools for numerically solving a wide range of differential equations, including stochastic jump processes. The solvers offer a means to define time-dependency on model parameters. \texttt{Optimization.jl} and \texttt{Turing.jl} offer optimization methods and Bayesian sampling methods, while \texttt{ModelingToolkit.jl} offers a means to symbolically define a dynamical system and perform structural analysis of the system. In R, the closest alternative to pySODM is \texttt{pomp} \citep{pomp}, which allows users to implement stochastic jump models by specifying its unobserved process and measurement components. \texttt{deSolve} \citep{Soetaert2010} can be used to solve ordinary and partial differential equations with time dependency on the model parameters. To further extend the functionalities of pySODM, the use of \texttt{scipy.integrate.solve\_ivp()} could be replaced with the JAX-based library \texttt{diffrax} \citep{Kidger2021}. Further, defining the system's observed states and the observation process in the model declaration may (slightly) simplify the construction of a posterior probability function used to calibrate the system to data.\\

\noindent In what follows, we demonstrate pySODM's applicability by building three models in the context of two case studies from different disciplines. In the first case study, the reaction rate of the esterification of D-glucose and lauric acid using an immobilized enzyme is calibrated to a series of eight batch experiments performed at different concentrations. Then, the calibrated enzyme kinetic model is used to forecast the yields when a tubular, continuous-flow reactor is packed with the immobilized enzyme. By applying the conservation of mass we arrive at a 1-dimensional partial differential equations (PDE) model which is subsequently discretized into coupled ODEs through the Method of Lines \citep{Sadiku2000}. The second case study is the calibration of a stochastic, age-stratified model for influenza to empirical data from the 2017-2018 influenza season in Belgium. In addition, we have used pySODM to build two \sars{} models for Belgium \citep{Alleman2021, Alleman2023a} and to implement a macro-economic Input-Output model to assess the economic impact of lockdown in Belgium \citep{Alleman2023b}, and to implement a coupled epidemiological-economic co-simulation for Belgian and Sweden (unpublished).\\

\noindent We omit excessive listings of pySODM's syntax in this work for two reasons. First, for the sake of brevity and clarity. Second, as pySODM is subject to continuous evolution, new insights may lead to alterations in the syntax. Consequently, a detailed exposition of the syntax in this article may risk rendering the content outdated. The syntax of the case studies presented here are available as tutorials on pySODM's documentation website: \url{https://twallema.github.io/pySODM}

\begin{figure}[h!]
    \centering
    \includegraphics[width=0.9\linewidth]{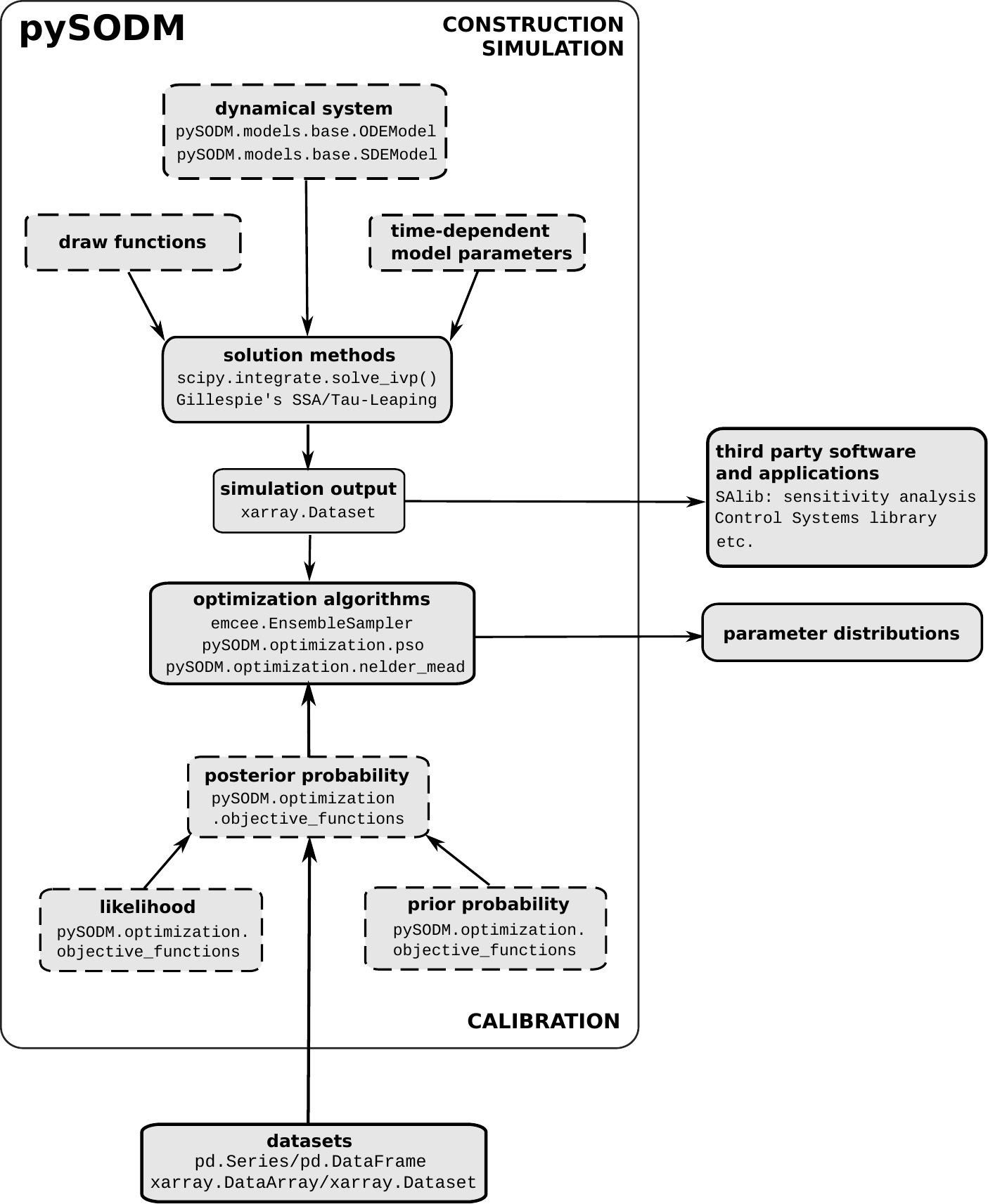}
    \caption{Conceptual representation depicting the structure of pySODM. Solid boxes depict the third-party implementations incorporated in pySODM, while the dashed boxes depict implementations provided by pySODM.} 
    \label{fig:conceptualization}
\end{figure}

\begin{landscape}
\begin{table}[!h]
    \centering
    \caption{An overview of pySODM's features.}
    \begin{tabular}{p{4cm}>{\raggedright\arraybackslash}p{13cm}}
        \toprule
        \textbf{Workflow} & \textbf{Features} \\ \midrule
        Construct a dynamical model & Implement coupled systems of ODEs \\
                             & States can be $n$-dimensional and of different sizes, allowing users to build models with subprocesses or implement PDEs by using the Method of Lines~\citep{Sadiku2000}.\\
                              & Allows $n$-dimensional model states to be labelled with coordinates and dimensions. \\
                              & Easy indexing, manipulating, saving, and piping to third-party software of model output by formatting simulation output as \texttt{xarray.Dataset}. \\
        \midrule
        Simulate the model & Deterministic (\texttt{scipy.integrate} \citep{Scipy2020}) or stochastic simulation (Gillespie's Stochastic Simulation Algorithm \citep{Gillespie1977} \& Tau-Leaping \citep{Gillespie2001}). \\
                           & Vary parameters dynamically using a generic and arbitrarily complex \textit{time-dependent parameter function}.\\
                           & Use \textit{draw functions} to perform repeated simulations for sensitivity analysis of model parameters. With \texttt{multiprocessing} support.\\
        \midrule
       Calibrate the model & Construct and maximize a posterior probability function.\\
                           & Automatic alignment of empirical data and model forecast over timesteps and dimensions.\\
                           & Nelder--Mead Simplex Optimization (local) and Particle Swarm Optimization (global) for point estimation of model parameters.\\
                           & Pipeline to and backend for \texttt{emcee.EnsembleSampler} to perform Bayesian inference of model parameters.\\
      \bottomrule
    \end{tabular}
    \label{tab:overview_features}
\end{table}
\end{landscape}

\section{Case studies}

\subsection{Enzymatic esterification in a 1D Packed-Bed Reactor}

\textbf{Introduction} Sugar fatty acid esters (SFAEs) are nonionic surfactants that play an important role in the food, detergent, agricultural, cosmetic and pharmaceutical industry. Because of several inherent merits and green character, the development of an enzymatic process is preferred over traditional chemical synthesis \citep{Zheng2015}. The combination of high conversion rates per volume unit, ease of scale-up by numbering-up, and inherent stability of lipases motivate the choice to synthesize SFAEs in continuous flow reactors packed with beads containing immobilized \textit{Candida Antartica} lipase B (CALB; brand name: Novozym 435). The esterification of D-glucose and lauric acid, performed in t-Butanol at 50 degrees Celcius and yielding glucose laurate ester and water as products, is used as a model reaction \citep{Flores2002}.\\

\ce{D-glucose + lauric acid
<=>[\ce{$50^{\circ}C$}][\ce{CALB}]
glucose laurate ester + H_2O
}

\subsubsection{Calibration of Intrinsic kinetics}

\noindent\textbf{Data collection} Multiple batch reactions were performed at different initial concentrations of D-glucose, lauric acid, and water \citep{Alleman2019}. Samples were withdrawn in threefold at regular intervals and analyzed for glucose laurate ester using an HPLC-MS. A detailed description of the lab protocol and the initial concentrations of the reactants is available in Appendix \ref{app:intrinsic_kinetics}.\\

\noindent\textbf{Batch reaction model} The rate equation used for this enzymatic esterification has nine parameters \citep{Flores2002}. Because the full rate equation of this reaction is convoluted, its calibration is typically performed in two steps. First, samples are withdrawn during the first minutes, this is referred to as an initial rate experiment. Under the assumption that no glucose laurate ester has yet been formed, a subset of six parameters can be calibrated. Second, to calibrate the three remaining parameters, the reaction is run until an equilibrium is reached, this is referred to as a full-time course experiment. We previously found that four parameters could be omitted from the rate equation \citep{Alleman2019}, a finding consistent with the work of Flores et al., 2002 \citep{Flores2002}. We consider the model reduction out-of-scope for this work and will simply calibrate the following reduced model,
\begin{eqnarray}
\dot{[S]} &=& - v,\nonumber \\
\dot{[A]} &=& - v,\nonumber \\
\dot{[Es]} &=& + v,\nonumber \\
\dot{[W]} &=& + v, 
\end{eqnarray}
where,
\begin{equation}\label{eqn:rate}
\frac{v}{[E]} = \frac{{V_f}/{K_S} ([S] [A] - (1/K_{eq}) [Es][W])}{[A] + R_{AS} [S] + R_{AW} [W] + R_{Es} [Es]},
\end{equation}\\

\noindent using data from three initial rate experiments, and five full-time-course experiments, starting with different concentrations of D-glucose, lauric acid, and water present (see Table \ref{tab:enzyme_kinetics_overview_concentrations}). For every measured concentration the relative error is available. $[E]$ is the enzyme concentration, constant and equal to 10 g/L. $[S]$ denotes the concentration of D-glucose, $[A]$ denotes the concentration of lauric acid, $[Es]$ denotes the concentration of glucose laurate ester, and $[W]$ denotes the concentration of water (in mM). $v$ is the reaction rate, expressed in millimolar per minute. The parameters $R_{AS}$, $R_{AW}$, and $R_{Es}$ (dimensionless) are interpreted as inhibitory constants due to their appearance in the denominator of the rate equation. $V_f/K_S$ is typically treated as one parameter. $V_f$ is the maximum rate of the forward reaction and is expressed in millimolar per minute and per gram biocatalyst while $K_S$ is a kinetic parameter expressed in millimolar. $K_{eq}$ is the equilibrium coefficient, expressed in millimolar, and determines if the reaction favors the reactants or the products.\\

\begin{figure}[h!]
    \centering
    \includegraphics[width=0.9\linewidth]{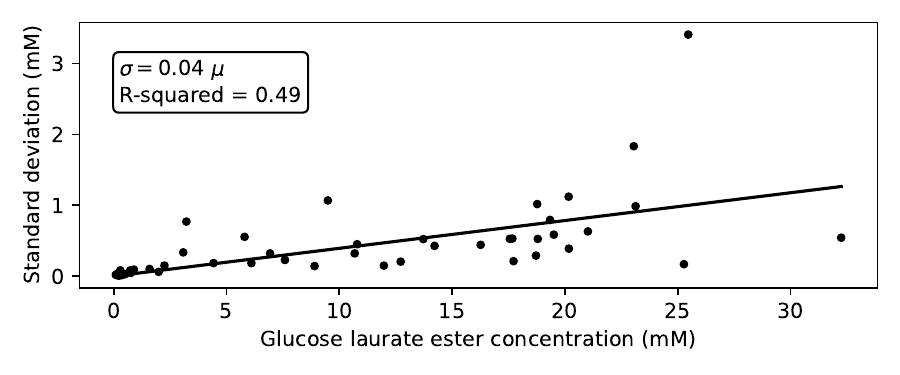}
    \caption{Relationship between magnitude of the measured glucose laurate ester concentration and the measurement's standard deviation.}
    \label{fig:mean-variance}
\end{figure}

\noindent\textbf{Model calibration} To perform an optimization of the model's parameters, an objective function measuring the mismatch of simulations and measurement data is needed. As an objective function, pySODM uses the parameter's posterior probability in light of the data, defined as \citep{Hartig2011},
\begin{equation}\label{eqn:posterior}
p (\theta \mid y) = \frac{p(y \mid \theta) p(\theta)}{p(y)},
\end{equation}
where $p (\theta \mid y)$ is the posterior probability, $p(y \mid \theta)$ is the likelihood, $ p(\theta)$ is the prior and $p(y)$, the probability of the data, is used as a normalization factor and can be neglected for all practical purposes. pySODM contains the necessary functions to align the model simulations and experimental observations and compute the logarithm of the posterior probability function. For each measured glucose laurate ester concentration, an error is available, we can thus analyze the mean-variance relationship to choose an appropriate likelihood function. In Figure \ref{fig:mean-variance}, the relationship between the magnitude of the measured glucose laurate ester concentration and the standard deviation is shown. The measurement standard deviation is heteroskedastic and equal to 4~\% of the measured concentration. We can use pySODM's Gaussian likelihood function, which is equal to a weighted sum of squares,
\begin{equation}
\log p(y \mid \bm{\theta}) = -0.5 \sum_{i=0}^N \sum_{j=0}^T \Bigg[ \frac{(y_{i,j} - \hat{y}_{i,j}(\bm{\theta}))^2}{\sigma_{i,j}^2} + \log (2 \pi \sigma_{i,j}^2) \Bigg],
\end{equation}
where the standard deviation of the measured concentration is equal to,
\begin{equation}
\sigma_{i,j} = 0.04\ y_{i,j},
\end{equation}
here $y_{i,j}$ is the glucose laurate ester concentration of the $j$th timestep of the $i$th experiment, $\hat{y}_{i,j}(\bm{\theta})$ is the glucose laurate ester concentration estimated using parameterset $\boldsymbol{\theta}$, and $\sigma_{i,j}$ is the standard deviation of the observations. Although it can be tempting to weigh the datapoints with the observed variability, this is generally not recommended when the number of replicates is low, as the deviation will vary considerably just by chance, potentially skewing the estimated model parameters \citep{Motulsky2005}. pySODM handles the bookkeeping related to computing the posterior probability using multiple datasets with different initial conditions. For each parameter, an uninformative (uniform) prior is used to bound the parameters to positive values. If the user has preconceptions about the values of parameters, pySODM supports the use of other prior probability distributions and L1/L2 prior regularisation \citep{hoerl1970}. First, pySODM's Particle Swarm Optimiser is used to scan the five-dimensional parameter space for a global maximum of the posterior probability (Eqn. \ref{eqn:posterior}). Then, the obtained estimate is perturbed uniformly by 10\%. The resulting perturbed values are used to start the affine-invariant ensemble sampler by Goodman and Weare, a Markov-Chain Monte-Carlo (MCMC) technique \citep{Goodman2010}. The sampler is run until the length of the chain is 50 times longer than the largest integrated autocorrelation. pySODM automatically produces diagnostic figures to follow up the sampling algorithm, such as the traceplot (Fig. \ref{fig:example_traceplot}) and autocorrelation plot (Fig. \ref{fig:example_autocorrelation}).\\

\begin{figure}[h!]
    \centering
    \includegraphics[width=1.02\linewidth]{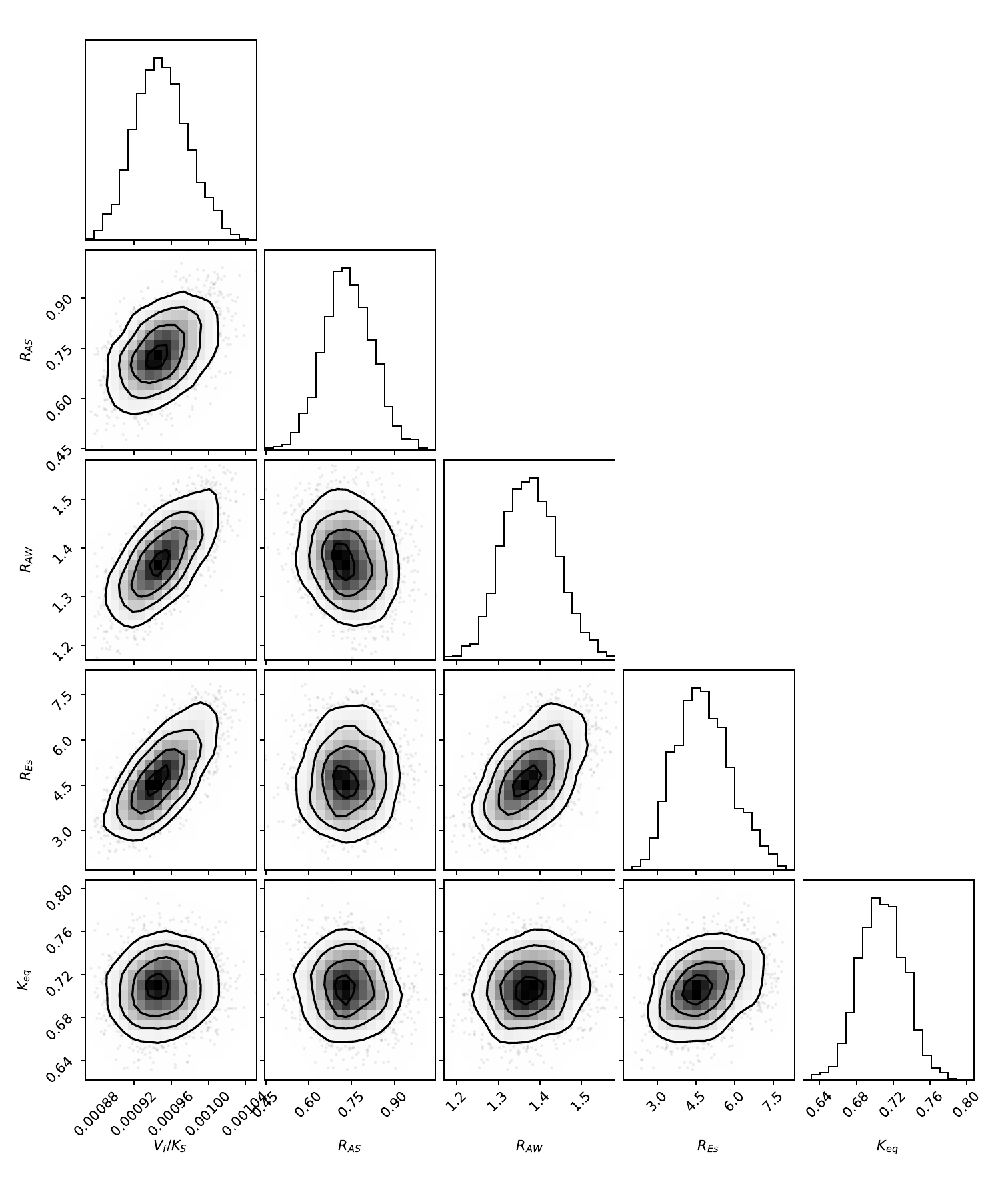}
    \caption{Two-dimensional visualizations (corner plot \citep{corner2016}) of the distributions of the rate equation's calibrated parameters.}
    \label{fig:enzyme_kinetics_corner}
\end{figure}

\begin{figure}[h!]
    \centering
    \includegraphics[width=0.9\linewidth]{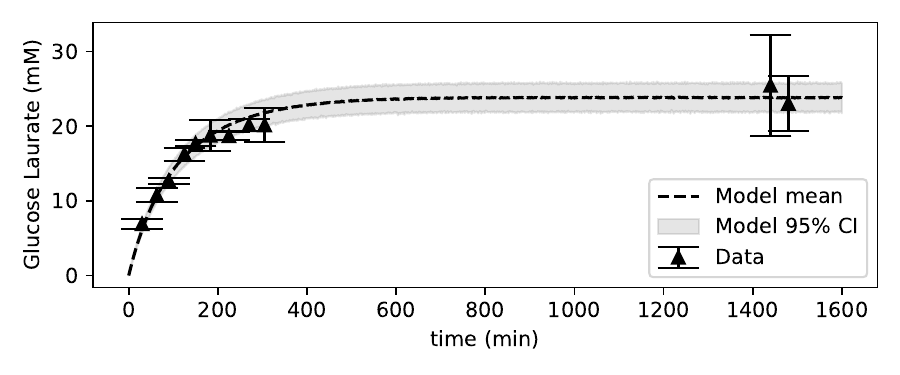}
    \caption{glucose laurate ester concentration as a function of time in a batch reaction initialized with 40 mM D-glucose, 121 mM lauric acid and 24 mM water. }
    \label{fig:enzyme_kinetics_fit1}
\end{figure}

\noindent\textbf{Results} In Figure \ref{fig:enzyme_kinetics_corner}, the two-dimensional distributions of the five calibrated parameters are visualised in a corner plot \citep{corner2016}. The equilibrium of this reaction is unfavourable and shifted towards the reactants, as indicated by an equilibrium constant of $K_{eq} = 0.70$ (95~\% CI: 0.64 - 0.76). It is thus likely that products will have to be removed during or in between reactions to attain higher yields. Figure \ref{fig:enzyme_kinetics_fit1} shows the goodness-of-fit over time for a reaction initialized with 40 mM D-glucose, 121 mM lauric acid and 24 mM water. After 24 hours, an equillibrium was reached and $24~\text{mM}$ (95\% CI: 22 mM - 26 mM) of glucose laurate ester was formed, meaning the reaction had a yield of $60\%$ (95\% CI: 55~\% - 65~\%). For this enzymatic reaction, higher acid-to-sugar ratios and lower initial water concentrations lead to the highest yields (see Figure \ref{fig:enzyme_kinetics_vary_AS}). A reaction initialized with 38 mM D-glucose, 465 mM lauric acid and 24 mM water reached a yield of $84~\%$ (95~\% CI: 79~\% - 89~\%). 

\subsubsection{Simulation of a Packed-Bed Reactor}

\begin{figure}[h!]
    \centering
    \includegraphics[width=0.80\linewidth]{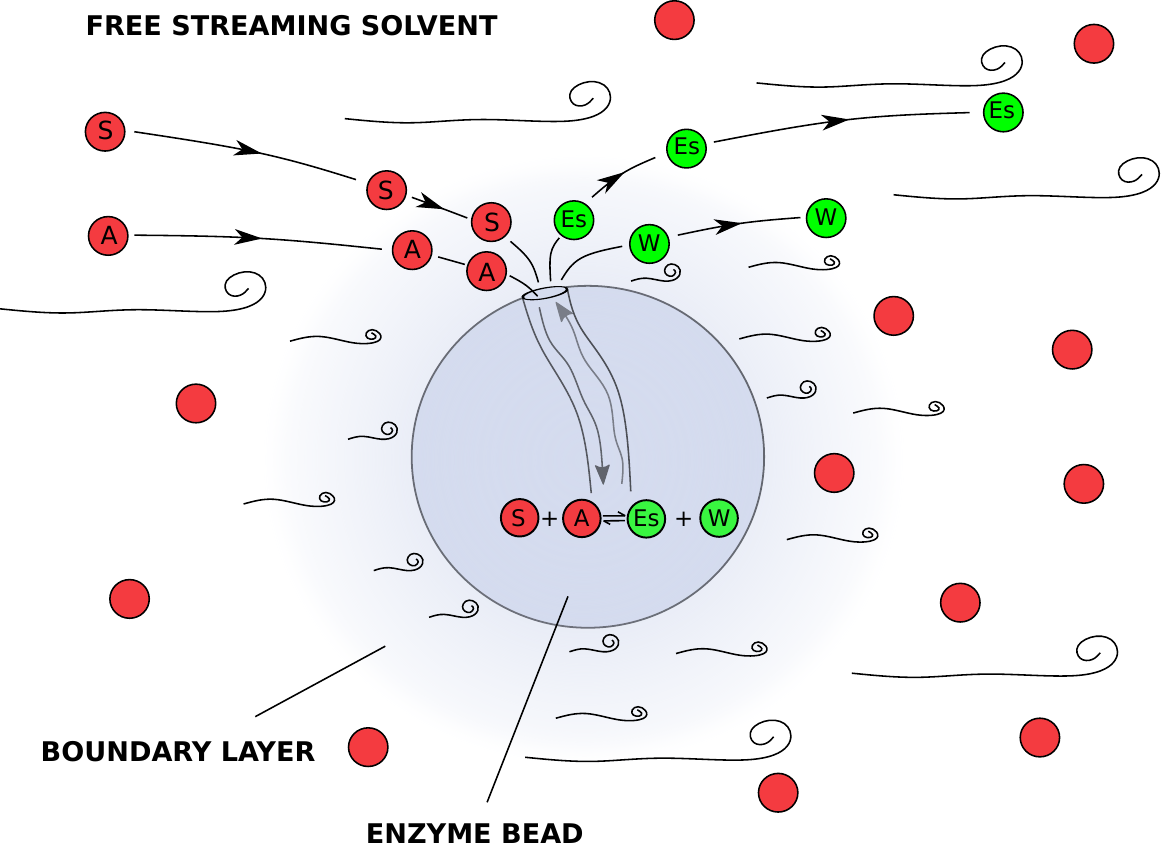}
    \caption{Mass transfer to and from the free streaming solvent to the enzyme located in the Novozym 435 enzyme beads, by means of external diffusion through the boundary layer and internal diffusion inside the beads macropores.}
    \label{fig:enzyme_kinetics_heterogeneous_catalyst}
\end{figure}

\noindent\textbf{Packed-bed reactor model} We now wish to use our calibrated rate equation to predict how the reaction progresses in a tubular, continuous flow reactor with an inner diameter of 2400 micrometer, packed with the beads containing our immobilized CALB (diameter 475 micrometer). In heterogeneous catalysis, mass transfer processes are often as important as the chemical reaction itself (Figure \ref{fig:enzyme_kinetics_heterogeneous_catalyst}). We introduce our reactants at the reactor inlet, where they stream freely between the enzymatic beads. The enzyme is located inside macropores in the beads and the reactants undergo two processes before they reach the enzyme. First, they diffuse through the boundary layer from the free-streaming solvent to the surface of the beads, this is called external diffusion. Second, they diffuse inside the pores of the catalyst to an enzyme molecule, this is called internal diffusion. Then, the esterification reaction takes place and the products move in opposite sequences to the free streaming solvent. As the pores inside the enzyme beads are much larger than the reactants and products, internal diffusion in our beads can luckily be neglected. By using the conservation of mass, we arrive at the following system of coupled PDEs (Appendix A),
\begin{eqnarray}\label{eq:packed_bed}
\dfrac{\partial C^{i,j}_{\mathrm{F}}}{\partial t} &=& \underbrace{D^i_{\mathrm{ax}} \dfrac{\partial^2 C^{i,j}_{\mathrm{F}}}{\partial x^2}}_\text{axial dispersion} - \underbrace{u \dfrac{\partial C^{i,j}_{\mathrm{F}}}{\partial x}}_\text{convection} + \underbrace{\dfrac{k_{\mathrm{L}} a^i}{\epsilon} (C^{i,j}_{\mathrm{S}}-C^{i,j}_{\mathrm{F}})}_\text{diffusion to catalyst},\nonumber \\
\dfrac{\partial C^{i,j}_{\mathrm{S}}}{\partial t} &=& \underbrace{\text{-} \dfrac{k_{\mathrm{L}} a^i}{(1\text{-}\epsilon)} (C^{i,j}_{\mathrm{S}}-C^{i,j}_{\mathrm{F}})}_\text{diffusion from catalyst} + \underbrace{\rho_{\mathrm{B}} \dfrac{v^i}{[E]}}_\text{reaction}\ ,
\end{eqnarray}
where $C_F^{i,j}$ represents the concentration of species $i$ at position $j$ along the reactor axis in the free streaming solvent and $C_S^{i,j}$ represents the concentration of species $i$ at position $j$ on the surface of the catalyst beads. $v_i/[E]$ is the intrinsic reaction rate calibrated previously (Equation \ref{eqn:rate}). $D_{ax}^i$ is the axial dispersion coefficient of species $i$, $k_L a^i$ is the mass transfer coefficient of species $i$ through the boundary layer, $\epsilon$ is the porosity of the packed bed, $\rho_B$ is the density of the Novozym 435 beads. All parameter values are listed in Table \ref{tab:enzyme_kinetics_overview_parameters}.\\

\noindent The system of PDEs can be converted to a system of ODEs by replacing the spatial derivatives with their respective first-order approximations following the Method of Lines \citep{Sadiku2000} (see Appendix \ref{app:packedbed_reactor}). At the reactor inlet, the concentration of all species is known and constant, and thus a Dirichlet boundary condition is used. At the outlet, a no-flux boundary condition is used. The model has two states, $C_F^{i,j}$ and $C_S^{i,j}$, each with two dimensions. The first is the chemical species: S, A, Es, and W. The second is the spatial position in the reactor, and there are $n_x$ spatial nodes. Thus, each state is a $(4 \times n_x)$ array, the states, and their labeled dimensions can easily be implemented in pySODM. The labels can then be used to ease handling of the output using the \texttt{xarray.Dataset} format (see Listing \ref{code:enzyme_kinetics_output}). In this example, all states have the same number of dimensions and thus shape, however, using pySODM it is possible to specify dimensions separately for all model states. This can be relevant in the context of dynamic transmission models for vector borne diseases, such as malaria, where we may be interested in the age bracket of the humans but not in the age of the mosquitoes.\\

\noindent\textbf{Results} To validate the model, two experiments were performed. A first experiment was performed using a reaction mixture containing 30 mM D-glucose, 60 mM lauric acid and 28 mM water. The reactants were pumped through the reactor at a constant flow rate of $0.2~\text{mL}.\text{min}^{\text{-}1}$, resulting in an average residence time of 13.5 minutes. After the outlet concentration had stabilized, three samples were withdrawn at the outlet. Then, the reactor was cut short by 0.10 m and the procedure above was repeated to obtain the reactant profile across the reactor length. Propagating the previously obtained uncertainty on the rate equation's parameters (Fig. \ref{fig:enzyme_kinetics_corner}) is easy using pySODM's \textit{draw functions}. These allow users to make changes to the model parameters between consecutive simulations. 100 simulations were performed, each with a new random sample drawn from the distributions of the previously obtained kinetic parameters (Listing \ref{code:draw_functions} in Appendix \ref{app:packedbed_reactor}). An additional dimension (\texttt{`draws'}) is automatically added to the simulation output to easily index the repeated simulations (Listing \ref{code:enzyme_kinetics_output_draws} in Appendix \ref{app:packedbed_reactor}). As seen in Figure \ref{fig:enzyme_kinetics_cut_reactor}, our packed-bed model does a good job of describing the laboratory data. Further, in a mere 13.5 minutes, the reaction has (almost) reached its equilibrium, much faster than in a batch reaction (Figure \ref{fig:enzyme_kinetics_fit1}).\\

\begin{figure}[h!]
    \centering
    \includegraphics[width=0.95\linewidth]{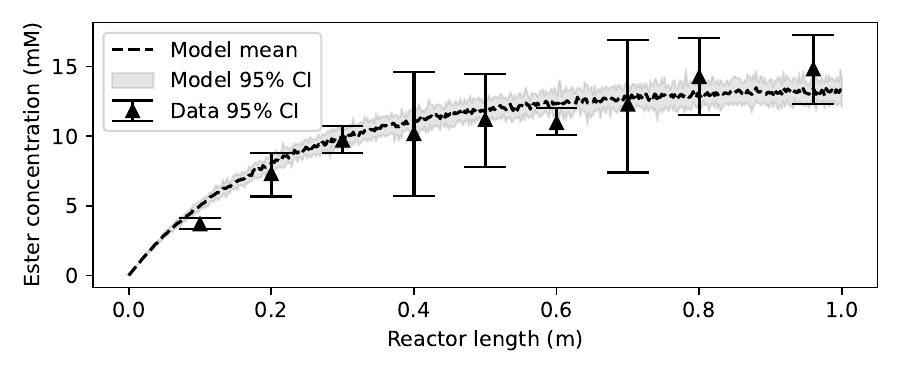}
    \caption{Glucose laurate ester concentration (mM) as a function of reactor length. A mixture of 30 mM D-glucose, 60 mM lauric acid and 28 mM water is fed at a flow rate $0.2~\text{mL}.\text{min}^{-1}$.}
    \label{fig:enzyme_kinetics_cut_reactor}
\end{figure}

\begin{figure}[h!]
    \centering
    \includegraphics[width=0.95\linewidth]{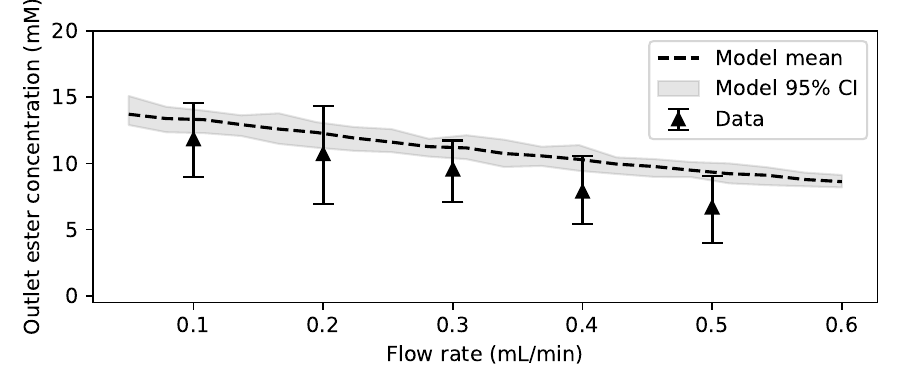}
    \caption{Glucose laurate ester concentration (mM) as a function of flow rate. A mixture of 30 mM D-glucose, 60 mM lauric acid and 18 mM water is fed through a reactor of $0.6~\text{m}$.}
    \label{fig:enzyme_kinetics_vary_flowrate}
\end{figure}

\noindent A second experiment was performed using a reaction mixture containing 30 mM D-glucose, 60 mM lauric acid, and 18 mM water. The reaction was initiated at a flow rate of $0.5~\text{mL}.\text{min}^{\text{-}1}$, which corresponded to a retention time of 5.4 minutes. The flow rate was then lowered to $0.1~\text{mL}.\text{min}^{\text{-}1}$ increments and samples were taken at the reactor outlet after a steady state was reached. As seen in Figure \ref{fig:enzyme_kinetics_vary_flowrate}, our model slightly overestimates the amount of product formed at high flow rates. This is likely caused by the tube's small inner diameter compared to the immobilized enzyme beads' diameter. The radial porosity profile of a packed bed is not uniform but oscillates. Near the container walls, the porosity is nearly $100~\%$ and the oscillations become smaller near the center of the packed bed (see Figure \ref{fig:packed_bed_radial_porosity}). The solvent likely channels faster through these regions of high porosity. In a tube with a bigger inner diameter the region of high porosity oscillations is smaller compared to the size of the packed bed and thus the observed effect should be smaller. The packed-bed reactor model can now be used to design a viable industrial process \textit{in silico}. A process of consecutive reaction-drying-reaction, where water is removed between two reaction stages, could drive up yields.

\subsection{A stochastic, age-stratified influenza model for the 2017-2018 season in Belgium}

\textbf{Introduction} Due to the annual recurrence of seasonal influenza, mathematical and computational models have been used widely in epidemiology to describe pandemic and seasonal transmission of influenza \citep{Brauer2019a}. In its yearly end-of-season report, the Belgian Institue for Public Health (Sciensano) publishes the weekly number of visits to general practitioners (GPs) with influenza-Like illness \citep{Bossuyt2018} (see Figure \ref{fig:influenza_data}). The 2017-2018 influenza season lasted 12 weeks and was of mild intensity \citep{Bossuyt2018}. In what follows, we build a (simple) stochastic dynamical transmission model for influenza and use pySODM to calibrate it directly to the age-stratified data.\\ 

\noindent\textbf{Transmission dynamics} We extend the classical Susceptible-Infectious-Recovered or SIR model of Kermack and McKendrick \citep{Kermack1927} by making two changes to the compartmental structure. First, an exposed state ($E$) is added to account for the latent phase between the moment of infection and the onset of infectiousness. Second, the infectious state ($I$) is split in three parts. Individuals may experience infectiousness prior to symptom onset ($I_{\text{pre}}$) \citep{Punpanich2012}. Then, after the onset of symptoms, not all infectious individuals will visit a GP and thus these individuals will not end up in the dataset. We include a state for individuals who are infectious but remain undetected ($I_{\text{ud}}$), and, we include a state for individuals who are infectious and go to a GP ($I_{\text{d}}$). All infectious individuals can transmit the disease. However, detected infectious individuals are assumed to only make 22~\% of the regular number of social contacts, corresponding to the fraction of contacts made at home. \\

\begin{figure}[h!]
    \centering
    \includegraphics[width=0.80\linewidth]{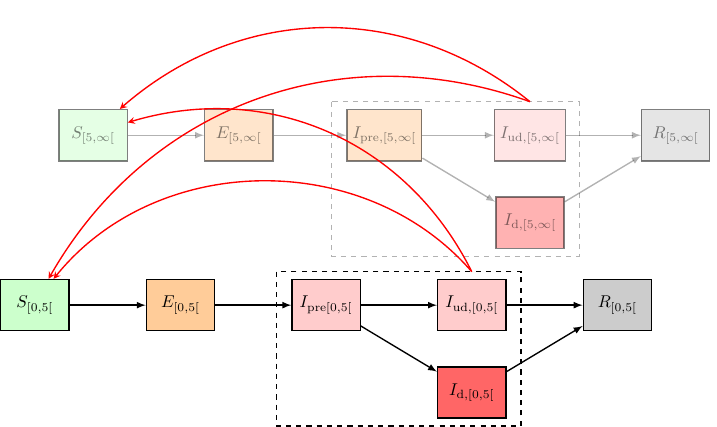}
    \caption{Example of an influenza model with $n=2$ age groups: $[0,5[$ and $[5, \infty[$. Infectious individuals have $n^2=4$ ways of infecting susceptibles. The model presented in this work has four age groups and thus there are 16 possible interactions.} 
    \label{fig:flowchart_influenza_model}
\end{figure}

\noindent Accounting for heterogeneity of the modeled population is an important aspect of disease modeling \citep{Brauer2019b}. The age of an individual determines the number of social contacts and the location where these contacts occur \citep{Mossong2008}, and the disease may progress differently for individuals of a different age \citep{Alleman2021}. Using pySODM to further extend compartmental dynamical transmission models with spatial entities and vaccinations is straightforward and was previously done in the context of the \sars{} epidemic \citep{Alleman2021,Alleman2023a}. We use pySODM's labeled $n$-dimensional states to split every disease compartment into four age groups: 0-5, 5-15, 15-65, and 65-120 years old. In this way, every disease state is now a one-dimensional vector containing four values.  A conceptual representation of the disease compartments, stratified in two age groups, is shown in Figure \ref{fig:flowchart_influenza_model}.\\

\noindent\textbf{Stochastic simulation} To simulate our model stochastically, we use the Tau-leaping method proposed by Gillespie \citep{Gillespie2001}, an approximation to the exact but computationally much more expensive Stochastic Simulation Algorithm \citep{Gillespie1977}. pySODM's stochastic model class requires users to define two functions: The first defines the rates of the transitions in the system (Eq. \ref{eq:transitions}), and the second defines how the transitions alter the system (Eq. \ref{eq:altering_states}). The dynamic transmission model depicted in Fig. \ref{fig:flowchart_influenza_model} has six possible transitions, 
\begin{eqnarray}\label{eq:transitions}
T^i &=& S^i + E^i + I_{\text{pre}}^i + I_{\text{ud}}^i + I_{\text{d}}^i + R^i, \nonumber\\
\mathcal{R}(S \rightarrow E)^i &=& \beta \sum_j N^{ij}(t) \dfrac{(I_{\text{pre}}^j + I_{\text{ud}}^j + 0.22 I_{\text{d}}^j)}{T^j}, \nonumber\\
\mathcal{R}(E \rightarrow I_{\text{pre}})^i &=& 1/\alpha, \nonumber \\
\mathcal{R}(I_{\text{pre}} \rightarrow I_{\text{ud}})^i &=& f_{\text{ud}}^i/\gamma, \nonumber \\
\mathcal{R}(I_{\text{pre}} \rightarrow I_{\text{d}})^i &=& (1-f_{\text{ud}}^i)/\gamma, \nonumber \\
\mathcal{R}(I_{\text{ud}} \rightarrow R)^i &=& 1/\delta, \nonumber \\
\mathcal{R}(I_d \rightarrow R)^i &=& 1/\delta, 
\end{eqnarray}

\noindent where the subscript $i$ refers to the aforementioned age groups. $T$ denotes the total population, $S$ denotes the number of individuals susceptible to the disease, $E$ denotes the number of exposed individuals, $I_{\text{pre}}$ denotes the number of presymptomatic infectious individuals, $I_{\text{ud}}$ denotes the number of infectious but undetected individuals and $I_{\text{d}}$ denotes the number of infectious individuals who visit the GP, $R$ denotes the number of removed individuals, either through death or recovery. The model has six parameters: $\alpha$, the length of the latent phase is equal to one day \citep{Punpanich2012}, $\beta$, the per-contact chance of influenza transmission or transmission coefficient (calibrated). $N^{ij}(t)$ is the square origin-destination matrix containing the number of social contacts in age group $i$ with individuals from age group $j$. Further, $f_{ud}^i$ is the fraction of undetected cases in age group $i$ (calibrated), $\gamma$ is the length of the presymptomatic infectious stage, equal to one day \citep{Punpanich2012}, $\delta$ is the duration of infectiousness and is equal to four days \citep{Punpanich2012}. Assuming the aforementioned transition rates (Eqs. \ref{eq:transitions}) from a generic state $X$ to a state $Y$ in age group $i$, denoted $\mathcal{R}(X \rightarrow Y)^i$, are constant over the interval $[t,t+\tau]$, the probability of a transition from a generic state $X$ to $Y$ happening in the interval $[t,t+\tau]$ is exponentially distributed, mathematically,
$$
\mathcal{P}(X \rightarrow Y)^i = 1 - e^{- \tau\mathcal{R}(X \rightarrow Y)^i}.
$$
The corresponding number of transitions $X \rightarrow Y$ in age class $i$ between time $t$ and $t+\tau$ are then obtained by drawing from a binomial distribution,
$$
\mathcal{N}(X \rightarrow Y)^i = \text{Binom}(\mathcal{P}(X \rightarrow Y)^i, X^i).
$$
The number of individuals in each of the compartments at time $t+\tau$ are then updated as follows,
\begin{eqnarray}\label{eq:altering_states}
S^i(t+\tau) &=& S^i(t) - \mathcal{N}(S \rightarrow E)^i, \nonumber \\
E^i(t+\tau) &=& E^i(t) + \mathcal{N}(S \rightarrow E)^i - \mathcal{N}(E \rightarrow I_{\text{pre}})^i, \nonumber\\
I_{\text{pre}}^i(t+\tau) &=& I_{\text{pre}}^i(t) + \mathcal{N}(E \rightarrow I_{\text{pre}})^i - \mathcal{N}(I_{\text{pre}} \rightarrow I_{\text{ud}})^i - \mathcal{N}(I_{\text{pre}} \rightarrow I_{\text{d}})^i, \nonumber\\
I_{\text{ud}}^i(t+\tau) &=& I_{\text{ud}}^i(t) + \mathcal{N}(I_{\text{pre}} \rightarrow I_{\text{ud}})^i - \mathcal{N}(I_{\text{ud}} \rightarrow R)^i, \nonumber \\
I_{\text{d}}^i(t+\tau) &=& I_{\text{d}}^i(t) + \mathcal{N}(I_{\text{pre}} \rightarrow I_{\text{d}})^i - \mathcal{N}(I_d \rightarrow R)^i, \nonumber \\
R^i(t+\tau) &=& R^i(t) + \mathcal{N}(I_{\text{ud}} \rightarrow R)^i + \mathcal{N}(I_\text{d} \rightarrow R)^i.
\end{eqnarray}
The daily number of GP visits (incidence) is computed as,
\begin{equation}
I_{\text{d, inc}}^i(t+\tau) = \mathcal{N}(I_{\text{pre}} \rightarrow I_{\text{d}})^i
\end{equation}
The leap value is determined by balancing the accuracy of the obtained results with the need for computational resources. A leap value of $\tau = 0.75~d$ was chosen. The basic reproduction number in age group $i$ of the equivalent deterministic model can be computed using the next-generation matrix approach introduced by Diekmann et al. \citep{Diekmann1990, Diekmann2009},
\begin{equation}\label{eq:basic_reproduction_number}
R_0^i = \beta \big(\gamma + f_{\text{ud}}^i \delta + 0.22 (1-f^i_{\text{ud}})\delta\big) \sum_j N^{ij}, 
\end{equation}
and the population basic reproduction number is computed as the weighted average over all age groups using demographic data \citep{Statbel}.\\

\noindent\textbf{Time-varying social contact function} Social contact is a key driver in the spread of respiratory pathogens and differs significantly between weekdays, weekends, and holidays \citep{Mossong2008}. Social contact matrices $N^{ij}$ were extracted separately for weekdays, weekends, and holidays using the Socrates data tool by Willem et al. \citep{Willem2020}. Only physical contacts were included and the number of contacts was integrated with the duration of the contact. During the 2017-2018 season, there were multiple holidays. To implement the necessary time-dependency of $N^{ij}(t)$, pySODM's \textit{time-dependent parameter functions} (TDPFs) can be used (see Listing \ref{code:tdpf} in the Supplementary Materials). In a TDPF, the user has access to all model states, model parameters, and any number of arbitrary parameters allowing the user to build arbitrarily complex functions.\\

\noindent\textbf{Model calibration} We desire to infer the basic reproduction number $R_0$ by calibrating the transmission coefficient, $\beta$, and the fraction of undetected cases, $\bm{f_{\text{ud}}}$. To this end, a posterior probability function must be set up. For count data, appropriate likelihood functions are the Poisson or Negative Binomial likelihood function, depending on the occurrence of overdispersion in the data. However, as only the average daily incidence of GP visits during a given week is available, it is not possible to estimate the relationship between the mean and variance of the data (as we previously did \citep{Alleman2023a}). As our likelihood function, we will assume the weekly case count is the result of seven counts, one per day, resulting from a Poisson observation process. We will thus use pySODM's built-in Poisson likelihood function, mathematically,
\begin{equation}
\log p(y \mid \bm{\theta}) = - \sum_{i=0}^N \sum_{j=0}^T \Bigg[ \hat{y}^{i,j}(\bm{\theta}) - y^{i,j} \log \hat{y}^{i,j}(\bm{\theta}) + y^{i,j}! \Bigg]
\end{equation}
where $y_{i,j}$ is the registered number of GP visits in age group $i$ of the $j$th datapoint, and $\hat{y}_{i,j}(\bm{\theta})$ is the predicted daily number of GP visits by age group $i$ on the date $t$ corresponding to the $j$th datapoint (proxied by model state $I_{\text{d, inc}}(t)$). Uninformative (uniform) priors are used to bound the parameters within physically plausible ranges. $\beta$ must be positive, while $\bm{f_{\text{ud}}}$ is bound between zero and one. To calibrate n-dimensional parameters using pySODM, such as the one-dimensional parameter $\bm{f_{\text{ud}}}$ in this example, no additional code is needed. Further, pySODM performs the necessary bookkeeping to align the age-stratified data with the age-stratified model output. The only condition is the dimensions and coordinates must match. Further, the user is free to simulate the influenza model with a larger number of age groups while calibrating to a dataset containing four age groups. An aggregation function can be defined to aggregate simulation output to the original four age groups. The use of aggregation functions has proven useful within the context of a spatially-explicit dynamic transmission model for \sars{} in Belgium \citep{Alleman2023a}, which we simulated at a finer spatial resolution than the available data.\\

\noindent We calibrate the presented model to an incrementally larger number of observations to assess the robustness of the calibration procedure. We start the calibration using only data until January 1st, 2018, and we then extend the number of available counts twice with one month, ending the calibration on February 1st, 2018, and March 1st, 2018. These moments are chosen to fall long before, right before, and after the influenza epidemic had peaked. To avoid bias during the calibration, the Particle Swarm Optimizer implemented in pySODM \citep{Kennedy1995}, which requires no user input, is first used to scan the five-dimensional parameter space. Then, the obtained maximum posterior probability is perturbated and the affine-invariant ensemble sampler \citep{Goodman2010} is run until the length of the chain is fifty times longer than the largest integrated autocorrelation.\\

\begin{figure}[h!]
    \centering
    \includegraphics[width=1.02\linewidth]{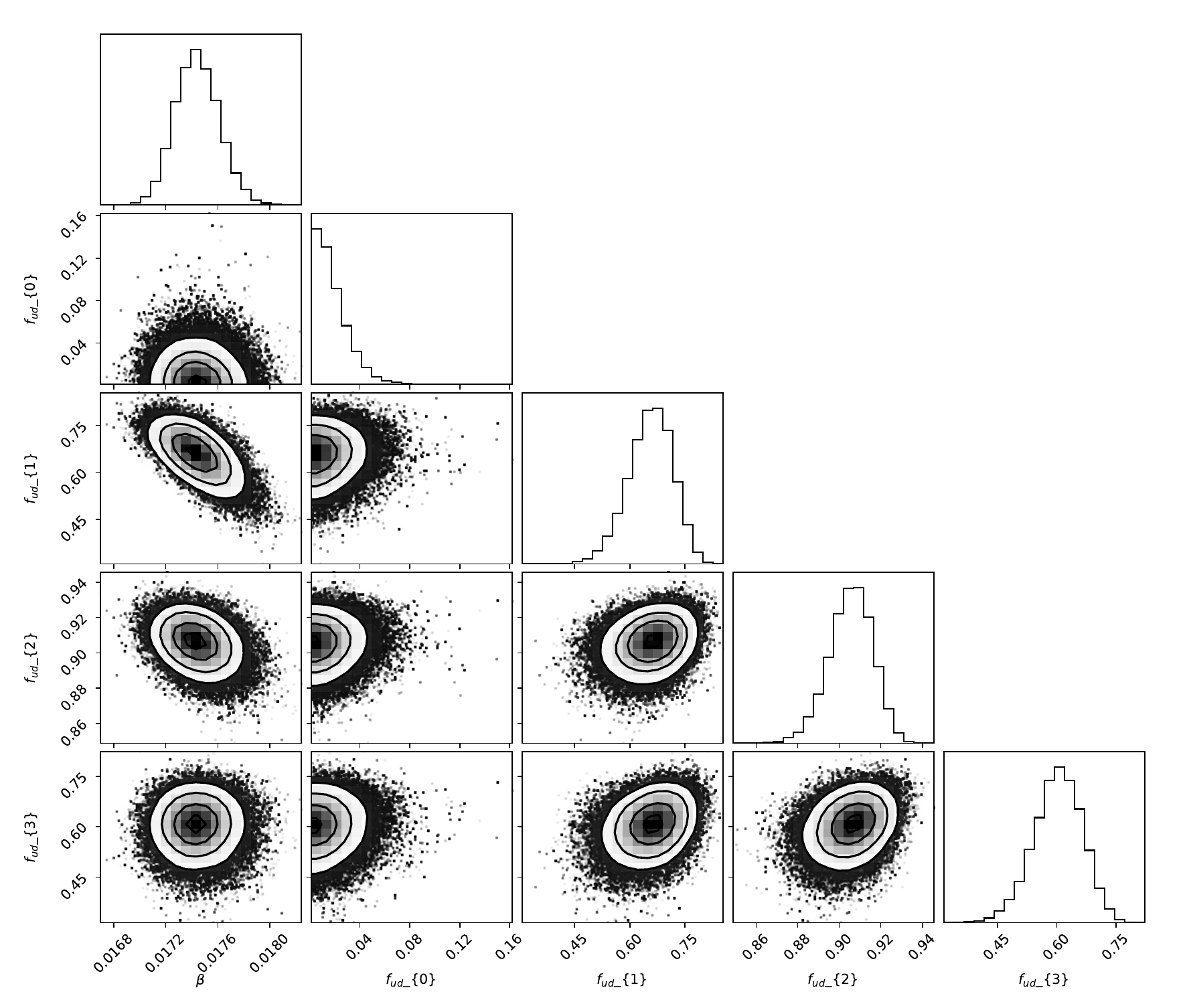}
    \caption{Two-dimensional visualizations (corner plot \citep{corner2016}) of the distributions of the influenza model's calibrated parameters.} 
    \label{fig:influenza_corner_March_1st}
\end{figure}

\noindent\textbf{Results} In Figure \ref{fig:influenza_corner_March_1st}, the distributions of the parameters $\beta$ and $\bm{f_{ud}}$, inferred using the largest dataset, are visualized. The optimal values of the fraction of undetected cases are $\bm{f_{ud}} = [0.01, 0.64, 0.90, 0.60]$. The undetected fraction is thus very small in children aged five years and below, then increases to 90~\% in individuals aged 15 to 65 years old, and finally decreases to 60~\% in the senior population. This finding is at least partly consistent with the findings of Dolk et. al (2021) \citep{Dolk2021}, who found the rate of GP consultations in the Netherlands to be three times higher in children aged five years and below. Some correlation between the infectivity ($\beta$) and the fraction of undetected cases in the age group of 5 to 15-year-olds ($f_{\text{ud}}\text{\_}\{1\}$, Fig. \ref{fig:influenza_corner_March_1st}) is visible. The population average basic reproduction number, calculated using Eq. \ref{eq:basic_reproduction_number} was equal to $R_0=1.95$ (95~\% CI: 1.91-1.98).\\

\begin{figure}[h!]
    \centering
    \includegraphics[width=0.8\linewidth]{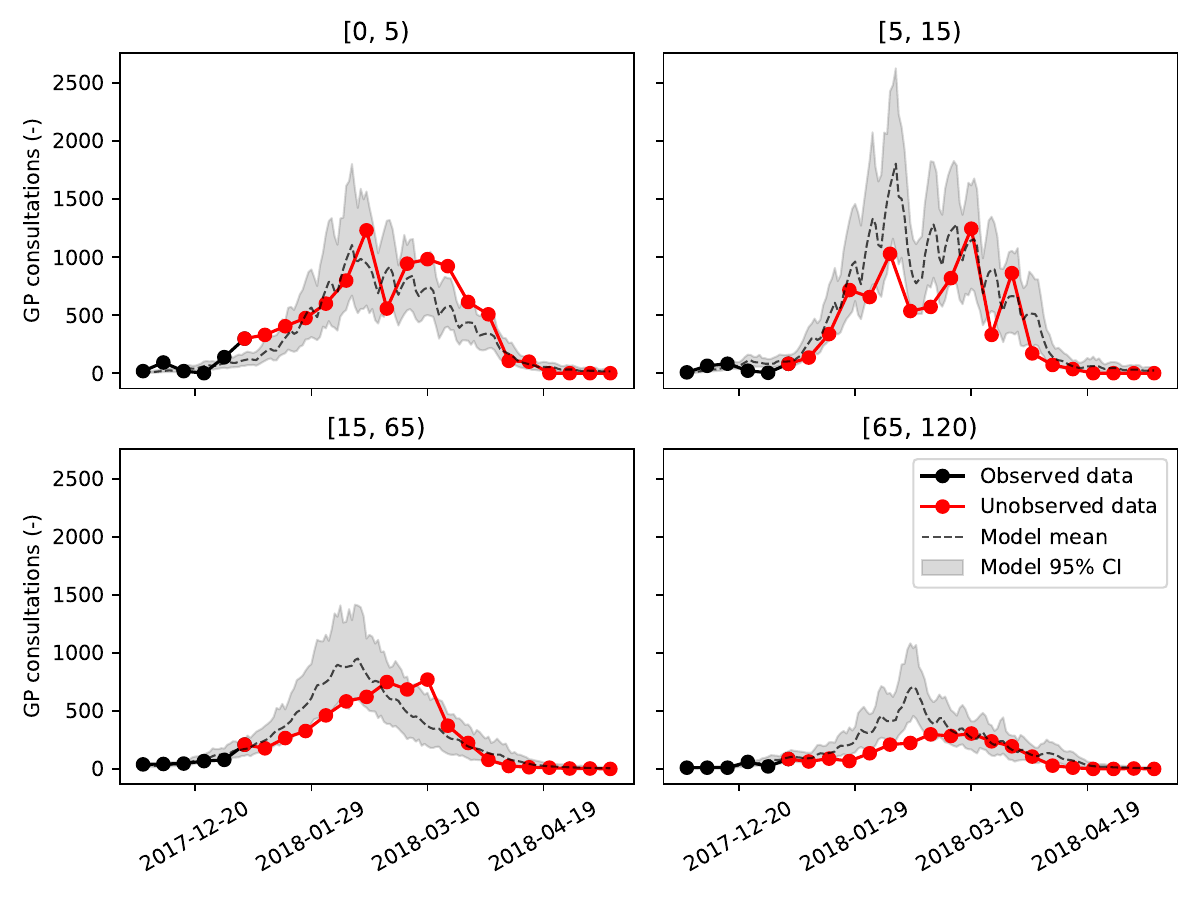}
    \caption{Modeled versus the simulated number of GP visits for influenza-like illness per 100.000 inhabitants. Calibration ended on January 1st, 2018.} 
    \label{fig:influenza_fit1}
\end{figure}

\begin{figure}[h!]
    \centering
    \includegraphics[width=0.85\linewidth]{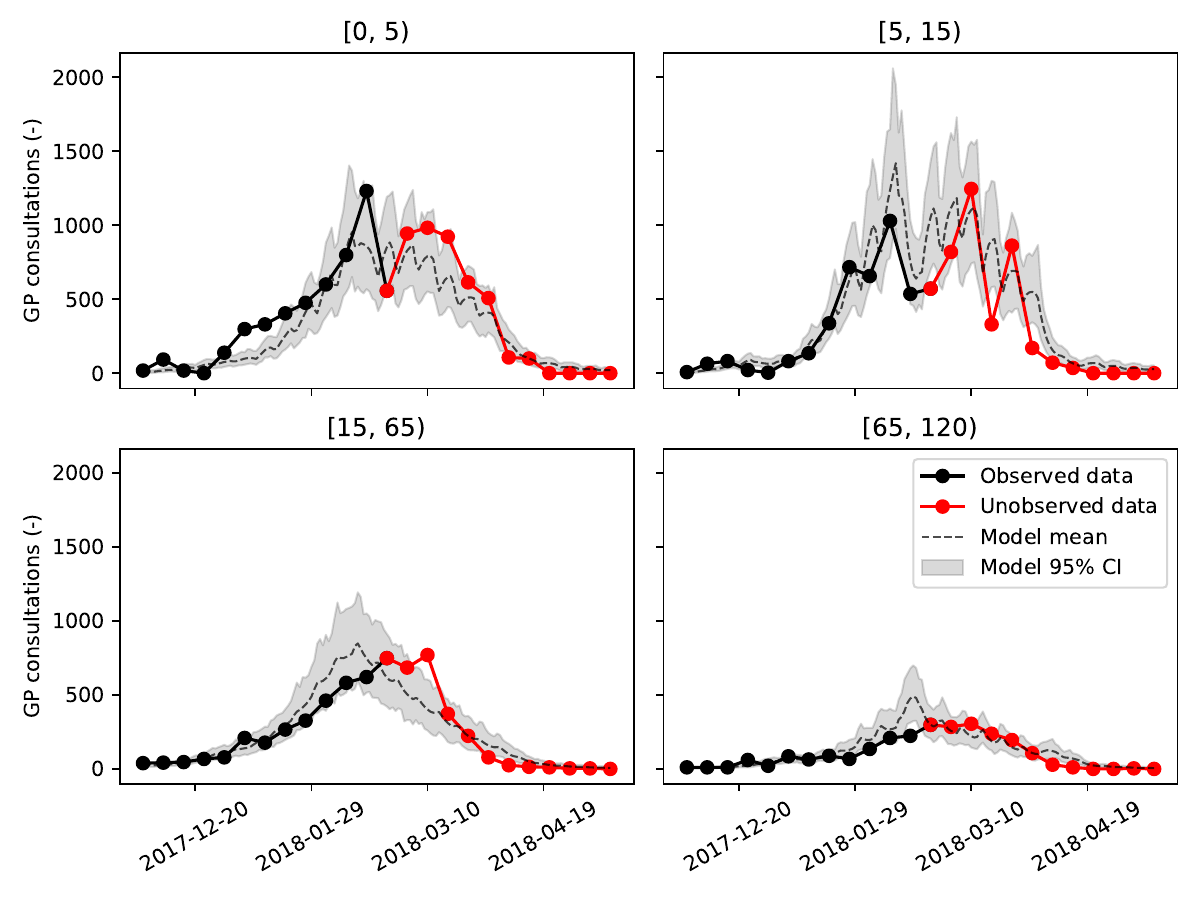}
    \caption{Modeled versus simulated number of GP visits for influenza-like illness per 100.000 inhabitants. Calibration ended on March 1st, 2018.} 
    \label{fig:influenza_fit3}
\end{figure}

\noindent Figures \ref{fig:influenza_fit1} \text{-} \ref{fig:influenza_fit3} show, for every age group and for the three calibrations performed, the result of 100 model trajectories and Poisson observational noise, plotted on top of the empirical data. Using the dataset ending on January 1st, 2018, the model is reasonably accurate and already provides a useful indication of the epidemic's peak magnitude and timing. The largest improvements in the model's accuracy are made for calibrations ending between January 1st, 2018 and, February 1st, 2018. The incidence of GP visits at the epidemic's peak is predicted with reasonable accuracy in all age groups. However, for the age groups $[15,65($ and $[65,120($, the predicted timing of the epidemic's peak falls two weeks prior to the observed epidemics's peak. For the age groups, $[0,5($ and $[5,15($, the timing of the epidemic's peak is adequate.\\

\noindent The results obtained using our simple model are encouraging but further research is needed before advising GPs and policy makers. First, by making the model spatially-explicit, we can include heterogeneity in the initial spread of Influenza, which will in turn render the predicted epidemic peaks more broad under the same number of social contacts. Second, including vaccines could likely further improve this model’s accuracy by lowering the peak incidence in the elderly population, as vaccine uptake was found to increase significantly in individuals above fifty years old \citep{Braeye2020}. Third, the consistency of the obtained parameter estimates, as well as the accuracy of the calibration procedure should be demonstrated across multiple influenza seasons. However, this is out of the scope as the aim of this work is merely to highlight our code's ability to speed up a modeling and simulation workflow.

\begin{figure}[h!]
    \centering
    \includegraphics[width=0.8\linewidth]{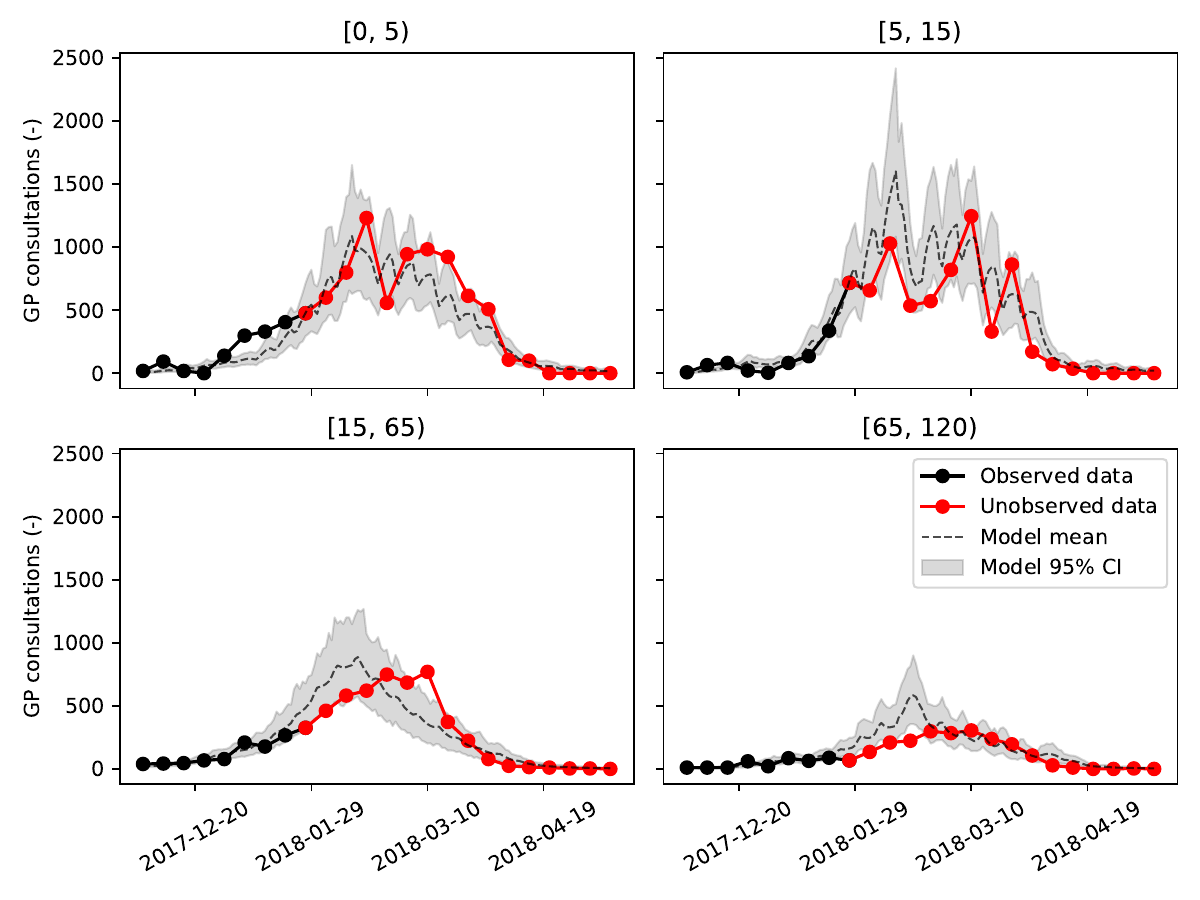}
    \caption{Modeled versus simulated number of GP visits for influenza-like illness per 100.000 inhabitants. Calibration ended on February 1st, 2018.} 
    \label{fig:influenza_fit2}
\end{figure}

\pagebreak
\section*{Conclusions}

In this work we introduced our generic framework to build, simulate and calibrate dynamical systems with labeled $n$-dimensional states in Python 3. pySODM integrates low-level interfaces for dynamical system simulation and calibration with the aim of speeding up commonly encountered workflows. Additionally, it offers generic functions to vary model parameters during the simulations, enables repeated simulations with parameter sampling, and includes a generic implementation of a posterior probability function for model and data alignment. We used our code to build a mathematical model based on partial differential equations for the enzymatic esterification of sugars and fatty acids in a packed-bed reactor, which could then be used for the \textit{in silico} design of a viable industrial process. We built an age-structured stochastic dynamic transmission model for influenza in Belgium and calibrated it to empirical data. Using limited data, our simple model was able to make a fairly accurate assessment of the future course of the epidemic. However, more research is needed before advising GPs and policy makers with the model. By building three models in the context of two case studies in different disciplines, reactor engineering and computational epidemiology, we demonstrated pySODM's applicability across scientific domains.

\pagebreak

\backmatter 

\bmhead{Supplementary information}

This work contains additional information on the case studies.

\bmhead{CRediT author statement}

\textbf{Tijs W. Alleman:} Conceptualisation, Methodology, Software, Investigation, Visualisation, Writing – original draft, Writing - Review \& Editing. \textbf{Christian Stevens:} Supervision, Funding acquisition, Resources \textbf{Jan M. Baetens:} Conceptualisation, Supervision, Funding acquisition, Project administration, Writing – Review \& Editing.

\bmhead{Acknowledgements}

TWA would like to acknowledge Prof. Ingmar Nopens' role in having Dr. Stijn Van Hoey and Dr. Joris Van den Bossche implement the first version of what became pySODM at the beginning of the \sars{} pandemic. TWA would also like to thank Dr. Jenna Vergeynst, Michiel Rollier and Wolf Demuynck for being the code's involuntary test subjects within the context of modeling \sars{} transmission in Belgium. This work was financially supported by \textit{Crelan}, the \textit{Ghent University Special Research Fund}, by the \textit{Research Foundation Flanders}, project numbers G0G2920 and 3G0G9820, and, by \textit{VZW 100 km Dodentocht Kadee} through the organisation of the 2020 100 km COVID-Challenge.

\bmhead{Conflict of interest} None declared.

\bmhead{Ethics approval} Not applicable.

\bmhead{Consent to participate} Not applicable.

\bmhead{Consent for publication} All authors have consented to publication of the manuscript in a peer-reviewed scientific journal, preceded by preprint publication in an open-access archive.

\pagebreak

\begin{appendices}

\section{Enzymatic esterification in a 1D Packed-Bed Reactor}

\subsection*{Calibration of Intrinsic Kinetics}\label{app:intrinsic_kinetics}

\textbf{Lab procedure} For each experiment a supersaturated solution of D-glucose and lauric acid in t-Butanol had to be prepared. First, as much water as possible had to be removed from the t-Butanol by means of 0.3 nm molecular sieves. Then, because of its low solubility in t-Butanol, a supersaturated solution of D-glucose was prepared by reflux boiling overnight. The maximum attainable concentration of D-glucose in t-Butanol at 50 Degrees Celcius is between 40 mM and 45 mM. Next, lauric acid was added and the mixture was transferred to a 50 mL flask suspended in an oil bath kept at 50 degrees Celcius. To start the reaction, 10 g/L of beads containing the enzyme were added to the mixture. The mixture was stirred with a magnetic stirrer throughout the reaction to avoid mass transfer limitations during the reaction course. Samples were withdrawn in threefold at regular intervals and analyzed for glucose laurate ester using an HPLC-MS.\\

\begin{table}[!h]
    \centering
    \caption{An overview of the initial concentrations of D-glucose, lauric acid and water (mM) at the start of the batch experiments. A full time course experiment is a batch reaction continued until an equilibrium is reached ($>24~\text{hr}$). An initial rate experiment is a batch reaction continued for 12 minutes, to determine the reaction rate in the abscence of product.}
    \begin{tabular}{p{2.5cm}>{\raggedright\arraybackslash}p{2cm}p{2cm}p{2cm}}
        \toprule
        \textbf{Experiment} & \textbf{D-glucose} & \textbf{lauric acid} & \textbf{water}\\ \midrule
      Full time course & 46.0 & 61.0 & 36.9 \\
                       & 40.5 & 121.5 & 24.3 \\
                       & 38.0 & 464.7 & 23.8 \\
                       & 30.0 & 60.2 & 304.0 \\
                       & 31.0 & 459.1 & 25.9 \\ \midrule
      Initial reaction rate & 20.1 & 20.5 & 28.2 \\   
                            & 40.0 & 40.0 & 36.0 \\   
                            & 44.0 & 150.0 & 24.0 \\                         
      \bottomrule
    \end{tabular}
    \label{tab:enzyme_kinetics_overview_concentrations}
\end{table}

\begin{figure}[h!]
    \centering
    \includegraphics[width=0.88\linewidth]{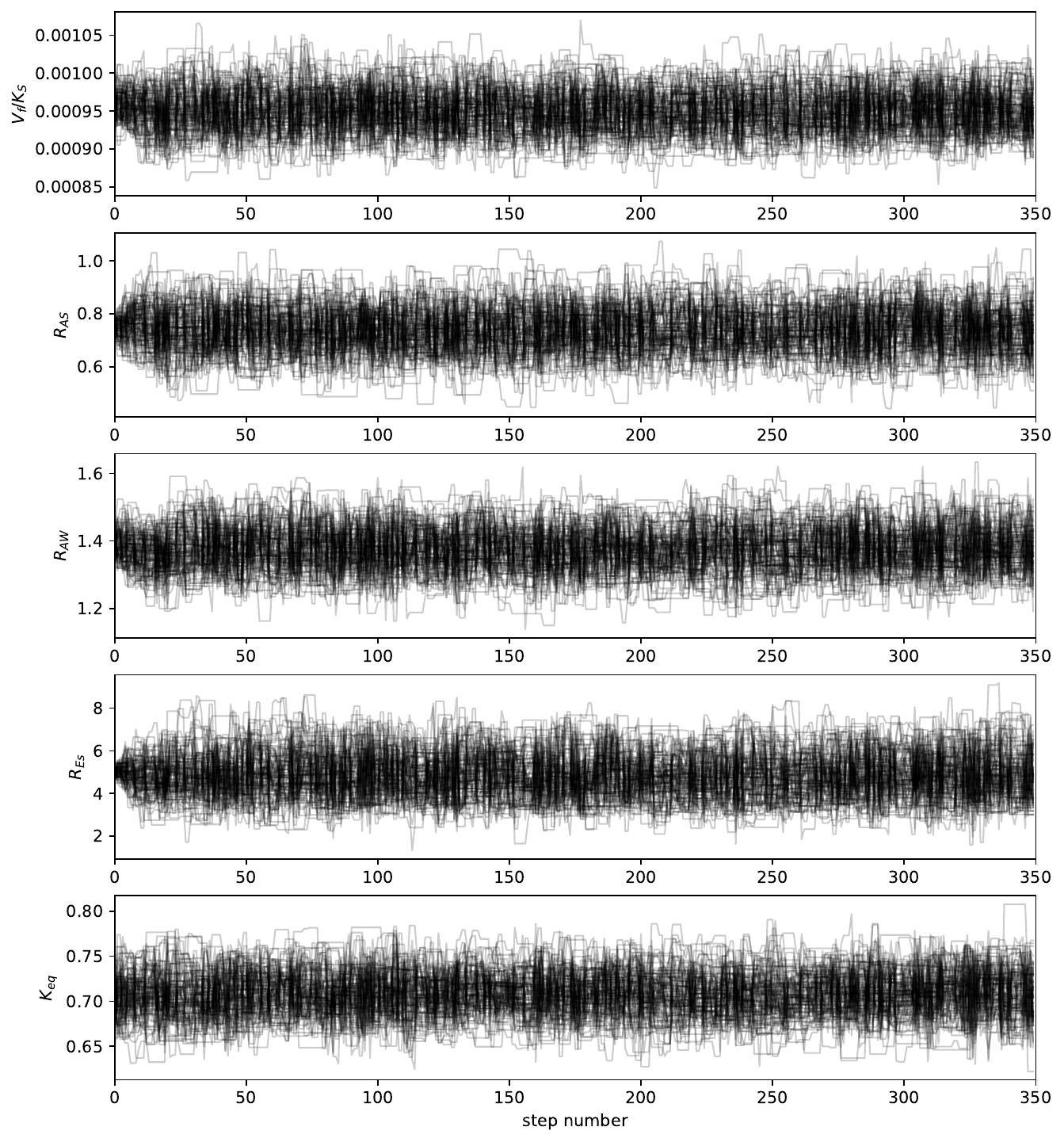}
    \caption{Markov chain traceplots (unthinned) of the intrinsic enzyme kinetics calibration. Diagnostic figures are automatically generated during MCMC sampling by pySODM.} 
    \label{fig:example_traceplot}
\end{figure}

\begin{figure}[h!]
    \centering
    \includegraphics[width=0.88\linewidth]{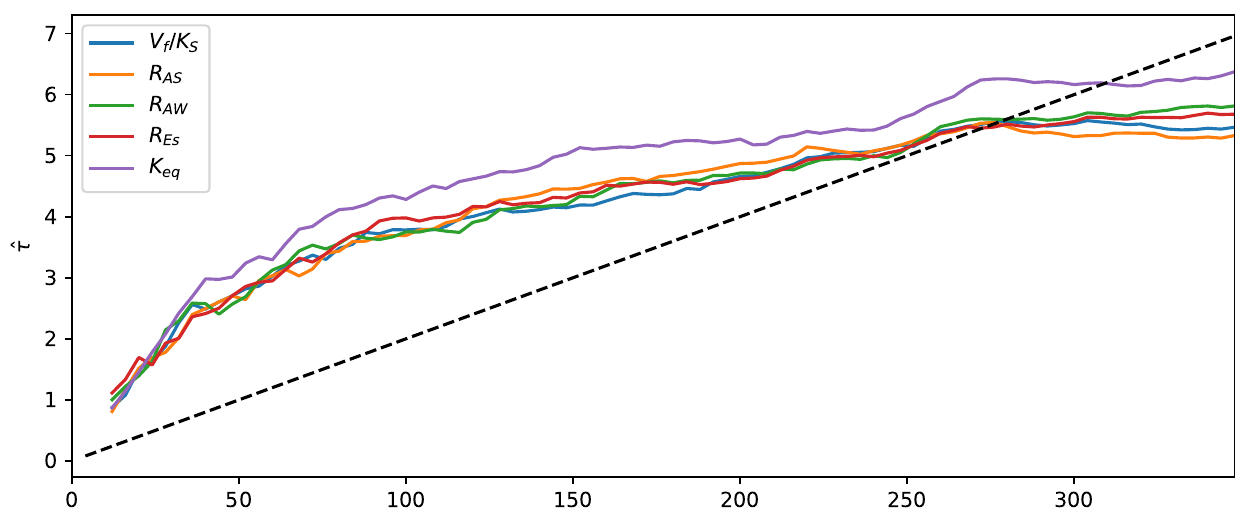}
    \caption{Autocorrelation of the Markov chains of the intrinsic enzyme kinetics model parameters. As a convergence criterion, the total number of steps must be 50 times greater than the largest autocorrelation. The dashed line represents the convergence criterion. Diagnostic figures are automatically generated during MCMC sampling by pySODM.} 
    \label{fig:example_autocorrelation}
\end{figure}

\begin{figure}[h!]
    \centering
    \includegraphics[width=0.80\linewidth]{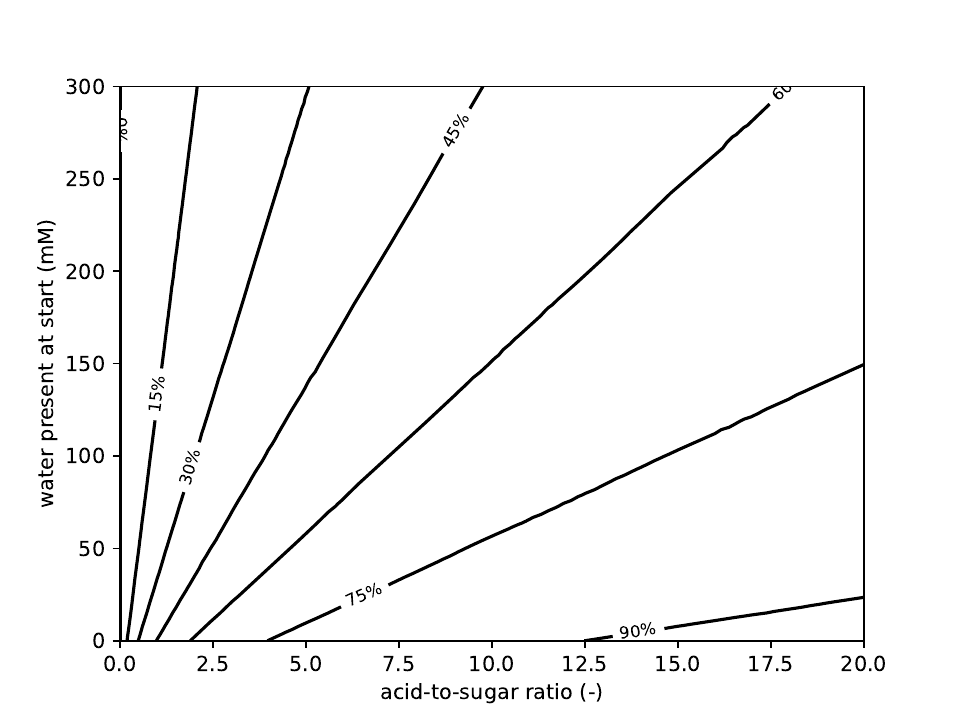}
    \caption{Reaction yield, defined as the percentage D-glucose conversion, on a 2D grid spanning the concentrations of D-glucose and lauric on the $x$-axis (given as the acid-to-sugar ratio with 40 mM of D-glucose used), and the water concentration on the $y$-axis. Lower initial water content and high acid-to-sugar ratios favor the formation of glucose laurate ester.} 
    \label{fig:enzyme_kinetics_vary_AS}
\end{figure}

\pagebreak 
\subsection*{Simulation of a Packed-Bed Reactor}\label{app:packedbed_reactor}

\noindent\textbf{Code listings}

\begin{lstlisting}[caption={Simulation output of the one-dimensional packed-bed reactor model, returned as an \texttt{xarray.Dataset} and containing the labeled $n$-dimensional states. },label={code:enzyme_kinetics_output},language=bash]
<xarray.Dataset>
Dimensions:  (time: 999, species: 4, x: 30)
Coordinates:
  * time     (time) int64 0 1 2 3 4 ... 996 997 998 999
  * species  (species) <U2 'S' 'A' 'Es' 'W'
  * x        (x) float64 0.0 0.0345 0.0690 0.103 ... 0.897 0.931 0.966 1.0
Data variables:
    C_F      (species, x, time) float64 30.0 30.0 30.0 ... 39.7 39.8 39.8
    C_S      (species, x, time) float64 30.0 30.0 30.0 ... 40.2 40.2 40.1

\end{lstlisting}

\begin{lstlisting}[caption={Example of a pySODM draw function. Draw functions allow users to make changes to the model parameters between consecutive simulations. Draw functions take the dictionary of model parameters (\texttt{param\_dict}) and an optional dictionary containing samples of model parameters (\texttt{samples\_dict}) as arguments. In the example below, we sample sets of model parameters from the posterior distributions obtained from MCMC sampling and assign them to the model's parameter dictionary. Draw functions are also usefull for sensitivity analysis, as we could sample model parameters from any distribution.},label={code:draw_functions},language=python]
def draw_fcn(param_dict, samples_dict):
    """
    A pySODM-compatible 'draw function' to sample enzyme kinetic parameters from `samples_dict` and assign them to `param_dict` between consecutive simulations.

    Input
    =====

    param_dict: dict
        Dictionary of model parameters

    samples_dict: dict
        Dictionary containing samples

    Output
    ======

    param_dict: dict
        Updated dictionary of model parameters
    """
    
    idx, param_dict['Vf_Ks'] = random.choice(list(enumerate(samples_dict['Vf_Ks'])))
    param_dict['R_AS'] = samples_dict['R_AS'][idx]
    param_dict['R_AW'] = samples_dict['R_AW'][idx]
    param_dict['R_Es'] = samples_dict['R_Es'][idx]
    param_dict['K_eq'] = samples_dict['K_eq'][idx]
    
    return param_dict

\end{lstlisting}

\begin{lstlisting}[caption={Simulation output containing 100 repeated simulations of the one-dimensional packed-bed reactor model, returned as an \texttt{xarray.Dataset}. An extra dimension \texttt{'draws'} has been added to the output as compared to Listing \ref{code:enzyme_kinetics_output} to accomodate the output of the repeated simulations.},label={code:enzyme_kinetics_output_draws},language=bash]
<xarray.Dataset>
Dimensions:  (time: 999, species: 4, x: 50, draws: 100)
Coordinates:
  * time     (time) int64 0 1 2 3 4 ... 996 997 998 999
  * species  (species) <U2 'S' 'A' 'Es' 'W'
  * x        (x) float64 0.0 0.0204 0.0408 0.0612 ... 0.959 0.980 1.0
Dimensions without coordinates: draws
Data variables:
    C_F      (draws, species, x, time) float64 30.0 30.0 ... 39.7 39.9
    C_S      (draws, species, x, time) float64 30.0 30.0 ... 40.4 40.4
\end{lstlisting}

\noindent\textbf{Conservation equations} The one-dimensional packed-bed reactor model assumes that all cross-sections are homogeneous and the radial movement and porosity distribution can be neglected. The packed bed is assumed to consist of two phases: (1) The bulk fluid in the interstices of the packed bed and (2) The enzyme beads surface where the reaction is assumed to take place. Dencic (2014) \citep{Dencic2014} concluded that for the Novozym 435 transesterification reaction of ethyl butyrate and 1-butanol, which is similar to the enzymatic reaction considered here, internal diffusion in the catalyst beads could be neglected. Neglecting internal diffusion allows to represent the system as if the reaction is happening at the surface of the catalyst pellet, drastically simplifying the model. The phases are separated by the mass transfer boundary layer around the spherical catalyst pellets (Figure \ref{fig:enzyme_kinetics_heterogeneous_catalyst}).\\

\begin{figure}[h!]
    \centering
    \includegraphics[width=0.80\linewidth]{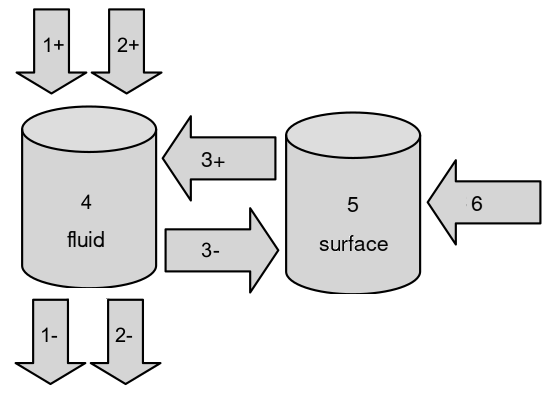}
    \caption{Schematic diagram of the ingoing and outgoing mass flows in an infinitisemal control volume of length $dx$ of the packed-bed reactor.} 
    \label{fig:enzyme_kinetics_reactor_slice_flowchart}
\end{figure}

\noindent First, let us focus on the fluid phase. A schematic diagram of a control volume of length $dx$ of both phases with the ingoing and outgoing mass flows is given in figure \ref{fig:enzyme_kinetics_reactor_slice_flowchart}. Species $i$ can enter control volume $j$ in three ways: through convective (1+) and diffusive (2+) transport and by diffusion from the catalyst surface through the liquid film (3+). Mass can leave the bulk liquid phase in three similar ways: through convective (1-) and diffusive (2-) transport, and by diffusion through the liquid film to the catalyst surface (3-). The mass of species $i$ entering control volume $j$ by convective means is equal to the convective flux,
\begin{equation}
(1+)\ \hspace{0.5cm} \underbrace{\epsilon u  C_F^{i,j}}_{\text{in}}\ ,
\end{equation}
where $u$ is the interstitial velocity of the packed bed and $U = \epsilon u$ is the superficial velocity or empty tube velocity of the packed bed (both in $\mathrm{m\cdot s^{\text{-1}}}$). $C_F^{i,j}$ denotes the bulk liquid concentration of species $i$ in control volume $j$ (mM). The mass leaving control volume $j$ by convection is equal to the mass entering control volume $j$ by convection plus the change over the control volume,
\begin{equation}
(1\text{-})\ \hspace{0.5cm} \underbrace{\epsilon u  C_F^{i,j} + \dfrac{\partial(\epsilon u C_F^{i,j})}{\partial x} dx}_{\text{out = in + change over $dx$}}\ .
\end{equation}
The difference in mass entering and leaving the control volume through convection is given by,
\begin{equation}
(1+)-(1\text{-})\ \hspace{0.5cm} \underbrace{\text{-}\epsilon u \dfrac{\partial C_F^{i,j}}{\partial x}dx}_{\text{in - out}}\ .
\label{eqn:convection}
\end{equation}
The diffusive transport term is derived in the same fashion as the convective term. Equation (2+) corresponds to Fick's law. This term is negative because mass is transferred diffusively from higher to lower concentrations. So, we get,
\begin{align}
(2+) \hspace{0.5cm} \phantom{=}& \text{-} \epsilon D_{\mathrm{ax}}^i  \dfrac{\partial C_F^{i,j}}{\partial x}\ ,\nonumber \\
(2\text{-}) \hspace{0.5cm} \phantom{=}& \text{-} \epsilon D_{\mathrm{ax}}^i  \dfrac{\partial C_F^{i,j}}{\partial x} + \dfrac{\partial }{\partial x}\Bigg( \text{-} \epsilon D_{\mathrm{ax}}^i \dfrac{\partial C_F^{i,j}}{\partial x} \Bigg) dx\ ,\nonumber \\ 
(2+)-(2\text{-}) \hspace{0.5cm} \phantom{=}&  \epsilon D_{\mathrm{ax}}^i \dfrac{\partial^2 C_F^{i,j}}{\partial x^2} dx\ ,
\label{eqn:diffusion}
\end{align}
where $D_{\mathrm{ax}}^i$ is the axial dispersion coefficient of species $i$ ($\mathrm{m^2 \cdot s^{\text{-}1}}$). The net mass diffusing through the boundary layer separating the bulk fluid from the catalyst surface is assumed to have a linear driving force. So it is assumed that mass transfer from the bulk liquid to the surface is lineary proportional to the bulk liquid concentration of species i and vice versa, 
\begin{align}
(3+) \hspace{0.5cm} \phantom{=}& k_{\mathrm{L}} a C_S^{i,j} dx\, \nonumber \\ 
(3\text{-}) \hspace{0.5cm} \phantom{=}& k_{\mathrm{L}} a C_F^{i,j} dx\, \nonumber \\
(3+)-(3\text{-}) \hspace{0.5cm} \phantom{=}&  k_{\mathrm{L}}a (C_S^{i,j} - C_F^{i,j}) dx\ ,
\label{eqn:transfer}
\end{align}
where $k_{\mathrm{L}}$ is the mass transfer coefficient ($\mathrm{m \cdot s^{\text{-}1}}$) and $a$ is the catalyst surface area ($\text{m}^{\text{-}1}$). The accumulation over the control volume becomes,
\begin{equation}
(4)\ \epsilon \dfrac{\partial C_F^{i,j}}{\partial t} dx\ .
\end{equation}\label{eqn:accumulation}

\noindent The general mass balance for the bulk fluid is computed by assuming that the accumulation in the bulk liquid phase is equal to the sum of the separate contributions,
\begin{equation}
\epsilon \dfrac{\partial C_F^{i,j}}{\partial t} dx = \epsilon D_{\mathrm{ax}}^i \dfrac{\partial^2 C_F^{i,j}}{\partial x^2} dx - \epsilon u \dfrac{\partial C_F^{i,j}}{\partial x}dx + k_{\mathrm{L}}a^i (C_S^{i,j} - C_F^{i,j}) dx\ ,
\end{equation}
and after dividing by $\epsilon$ and $dx$,
\begin{equation}
\dfrac{\partial C_F^{i,j}}{\partial t}  =  D_{\mathrm{ax}}^i \dfrac{\partial^2 C_F^{i,j}}{\partial x^2}  - u \dfrac{\partial C_F^{i,j}}{\partial x} + \dfrac{k_{\mathrm{L}}a^i}{\epsilon} (C_S^{i,j} - C_F^{i,j})\ .
\label{eqn:conservationLiquid}
\end{equation}
Similar to equation A6, accumulation at the catalyst surface of the control volume is equal to,
\begin{equation}
(5)\ (1-\epsilon) \dfrac{\partial C_S^{i,j}}{\partial t} dx\ ,
\end{equation}
where $C_S^{i,j}$ is the catalyst surface concentration of species $i$ (mM). The mass of species species $i$ formed or used by the enzyme is,
\begin{equation}
(6)\ (1 - \epsilon)\rho_{\mathrm{B}} \dfrac{v^i}{[E]_{\mathrm{t}}} dx\ ,
\end{equation}
where $\rho_{\mathrm{B}}$ is the catalyst bulk density ($\mathrm{g \cdot l^{\text{-}1}}$) and $v_{\mathrm{S,i}}/[E]$ the enzymatic reaction rate (Equation \ref{eqn:rate}) in units $\mathrm{mmol \cdot g^{\text{-}1} \cdot s^{\text{-}1}}$. The net mass diffusing through the liquid film is equal to (3-) - (3+). The resulting conservation equation for the catalyst surface is,
\begin{equation}
\dfrac{\partial C_S^{i,j}}{\partial t} = \text{-} \dfrac{k_{\mathrm{L}}a^i}{(1-\epsilon)} (C_S^{i,j} - C_F^{i,j}) + \rho_{\mathrm{B}} \dfrac{v^i}{[E]}\ .
\label{eqn:conservationSolid}
\end{equation}
The system of equations governing transport and chemical reaction in a one-dimensional, continuous flow, packed-bed tubular reactor is,
\begin{eqnarray}\label{eqn:packed_bed_app}
\dfrac{\partial C_F^{i,j}}{\partial t} &=& \underbrace{D_{\mathrm{ax}}^i\dfrac{\partial^2 C_F^{i,j}}{\partial x^2}}_\text{axial dispersion} - \underbrace{u \dfrac{\partial C_F^{i,j}}{\partial x}}_\text{convection} + \underbrace{\dfrac{k_{\mathrm{L}} a^i}{\epsilon} (C_S^{i,j}-C_F^{i,j})}_\text{diffusion to catalyst}, \nonumber \\
\dfrac{\partial C_S^{i,j}}{\partial t} &=& \text{-} \underbrace{\dfrac{k_{\mathrm{L}} a^i}{(1\text{-}\epsilon)} (C_S^{i,j}-C_F^{i,j})}_\text{diffusion to catalyst} + \underbrace{\rho_{\mathrm{B}} \dfrac{v^i}{[E]}}_\text{reaction}\ .
\end{eqnarray}

\noindent \textbf{Method of Lines} We use the Method of Lines \citep{Sadiku2000} to implement these equations in the pySODM framework. This method involves discretizing only the spatial derivatives to obtain a system of ODEs. We replace the spatial derivatives with their respective first order approximations. It is common practice to treat the convective term explictly while the diffusive term is treated implicitly.
$$
\begin{cases}
\dfrac{\partial C^{i,j}_F}{\partial x} &\approx \dfrac{C_F^{i, j+1} - C_F^{i, j}}{\Delta x},\\

\dfrac{\partial^2 C^{i,j}_F}{\partial x^2} &\approx \dfrac{C_F^{i, j-1} - 2 C_F^{i, j} + C_F^{i, j+1} }{\Delta x^2}.\\
\end{cases}
$$
Substituting these expressions in Equation \ref{eqn:packed_bed_app} we get,
$$
\begin{cases}
\dfrac{d C^{i,j}_{\mathrm{F}}}{d t} &= D^i_{\mathrm{ax}} \dfrac{C_F^{i, j-1} - 2 C_F^{i, j} + C_F^{i, j+1} }{\Delta x^2} - u \dfrac{C_F^{i, j+1} - C_F^{i, j}}{\Delta x} + \dfrac{k_{\mathrm{L}} a^i}{\epsilon} (C^{i,j}_{\mathrm{S}}-C^{i,j}_{\mathrm{F}}),\nonumber \\
\dfrac{d C^{i,j}_{\mathrm{S}}}{d t} &= \text{-} \dfrac{k_{\mathrm{L}} a^i}{(1\text{-}\epsilon)} (C^{i,j}_{\mathrm{S}}-C^{i,j}_{\mathrm{F}}) + \rho_{\mathrm{B}} \dfrac{v^i}{[E]_{\mathrm{t}}}.
\end{cases}
$$
All that is left is to consider what happens at the inlet and outlet boundaries. At the inlet ($j=0$), we will assume that the species concentration in both the liquid and at the catalyst surface are equal to fixed inlet concentrations $c^i$ provided by the user. Mathematically,
$$
\begin{cases}
C_F^{i,0} &= c^i,\\
C_S^{i,0} &= C_F^{i,0},
\end{cases}
$$
and thus,
$$
\begin{cases}
\dfrac{d C^{i,0}_{\mathrm{F}}}{d t} &= 0,\\
\dfrac{d C^{i,0}_{\mathrm{S}}}{d t} &= 0.
\end{cases}
$$
At the outlet ($j=N$), a problem arises as $C_F^{i, N+1}$ is needed to approximate our spatial derivatives and this node is outside our reactor domain. We can overcome this by treating our outlet as a no-flux boundary,
$$
\dfrac{d C^{i,N}_{\mathrm{F}}}{d t} = 0.
$$
Approximating the dervative in the LHS with a central finite difference approximation,
$$
\dfrac{C^{i,N+1} - C^{i,N-1}}{2 \Delta x} = 0,
$$
we can thus substitute $C_F^{i, N+1} = C_F^{i,N-1}$ at the reactor outlet.

\pagebreak
\noindent\textbf{Radial voidage distribution}

\begin{figure}[h!]
    \centering
    \includegraphics[width=0.80\linewidth]{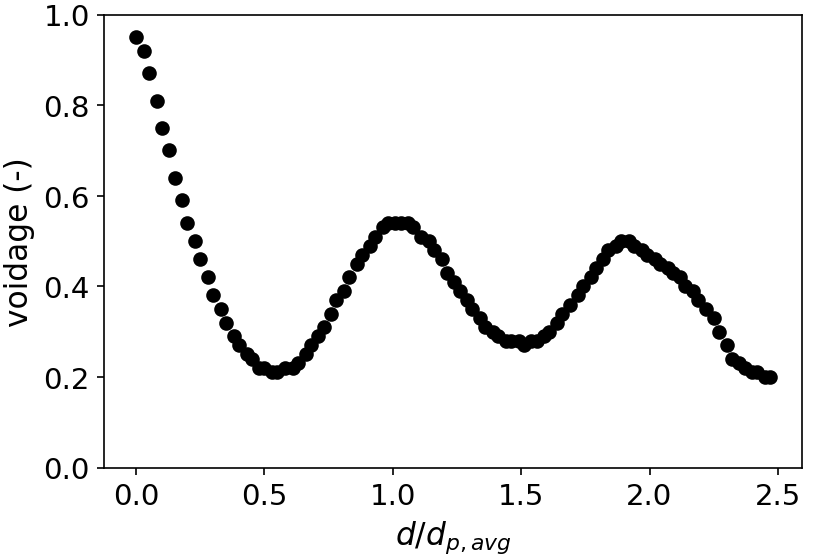}
    \caption{Simulated radial voidage inside a tube with a diameter of $2400~\micro m$, packed with 1000 enzyme beads with a diameter of $475\pm74~\micro m$. Near the container wall, the voidage is nearly $100~\%$. Simulated using BPG by Partopour and Dixon \citep{Partopour2017}.} 
    \label{fig:packed_bed_radial_porosity}
\end{figure}

\begin{landscape}
\thispagestyle{empty}
\begin{table}
    \centering
    \caption{\small{Overview of parameters used in the simulation of the packed-bed reactor.}}
    \begin{tabular}{p{1cm}p{6cm}p{5cm}p{6.5cm}}
        \toprule
        \textbf{Symbol} & \textbf{Parameter} & \textbf{Value(s)} & \textbf{Computation} \\ \midrule
        $l$ & Reactor length & $0.6~\mathrm{m}$ or $1~\mathrm{m}$ \\
        $n_x$ & Number of spatial nodes & $50$ \\
        $d_r$ & Reactor diameter & $0.0024~\mathrm{m}$ \\
        $d_p$ & Enzyme bead diameter & $0.0004755~\mathrm{m}$ \\
        $\rho_B$ & Catalyst density & $545~\mathrm{kg} . \mathrm{m}^{-3}$ \\
        $\mu$ & t-Butanol dynamic viscosity & $3.35*10^{-3}~\mathrm{Pa} . \mathrm{s}$ \\
        $Q$ & Flow rate & $0.2-0.6~\mathrm{mL} . \mathrm{min}$ \\
        $\rho_F$ & t-Butanol density & $775~\mathrm{kg} . \mathrm{m}^{-3}$ \\
        \midrule
        $u$ & Fluid velocity & $0.0017-0.0051~\mathrm{m}.\mathrm{s}^{-1}$ & $Q/ (\epsilon A)$ \\
        $Re$ & Reynolds number & $0.14-0.43$ & $\epsilon u \rho_F  d_p/ ( \mu (1-\epsilon))$ \\
        $a$ & Catalyst surface area & $7124~\mathrm{m}^{-1}$ & $6(1-\epsilon)/d_p$\\
        $\mathbf{D_{AB}}$ & Molecular diffusion coefficient in t-Butanol & $[0.35, 0.23, 0.20, 1.39]*10^{-6}~ \mathrm{m}^2 . \mathrm{s}^{-1}$ & Group contribution method detailed in \citep{Li1997, Schotte1992} \\
        $\mathbf{k_L}$ & Mass transfer coefficient & $[2.17, 2.09, 2.06, 2.90]*10^{-6}~\mathrm{m} . \mathrm{s}^{-1}$ & $k_L = 0.7 D_{AB} + \epsilon u d_p/(0.18+0.008 Re^{0.59})$; \citep{Carrara2003}\\
        $\mathbf{D_{ax}}$ & Axial dispersion coefficient &  $[8.0, 6.1, 5.4, 20.1]*10^{-6}~\mathrm{m}^2 . \mathrm{s}^{-1}$ & $D_{ax} = (1.09/100)*(D_{AB}/d_p)^{(2/3)}*\epsilon u^{(1/3)}$; \citep{Rastegar2017}\\
        $\epsilon$ & Porosity & 0.43 & $\epsilon = 0.39 + 1.74/(d_r/d_p+1.140)^2$; \citep{Benyahia2005}\\
        \bottomrule
    \end{tabular}
    \label{tab:enzyme_kinetics_overview_parameters}
\end{table}
\end{landscape}

\section{A stochastic, age-stratified influenza model for the 2017-2018 season in Belgium}\label{app:influenza}

\begin{lstlisting}[caption={An example of a (simplified) contact function altering the number of social contacts $N^{ij}$ during holidays. Time-dependent parameter functions (TDPFs) allow users to vary model parameters during the course of a single simulation. They take the simulation timestep (\texttt{t}), the dictionary of model states (\texttt{states}), and the value of the parameter to be changed (\texttt{param}) as obligatory inputs. In addition, TDPFs can take any number of user-defined parameters as inputs, allowing the user to build arbitrarily complex functions.},label={code:tdpf},language=python]
def contact_function(t, states, param, N_noholiday, N_holiday):
    """
    A pySODM-compatible `time-dependent parameter function` to vary social contacts during holidays

    Input
    =====

    t: timestamp
        Current date in simulation

    states: dict
        Dictionary containing model states at previous timestep

    param: dict
        Dictionary of model parameters

    N_noholiday: np.ndarray
        Contact matrix during non-holidays

    N_holiday: np.ndarray
        Contact matrix during holidays

    Output
    ======

    N(t): np.ndarray
        Contact matrix at date `t`
    """
    
    if t <= pd.Timestamp('2017-12-20'):
        return N_noholiday
    elif pd.Timestamp('2017-12-20') < t <= pd.Timestamp('2018-01-05'):
        return N_holiday
    elif ...
\end{lstlisting}

\begin{figure}[h!]
    \centering
    \includegraphics[width=1\linewidth]{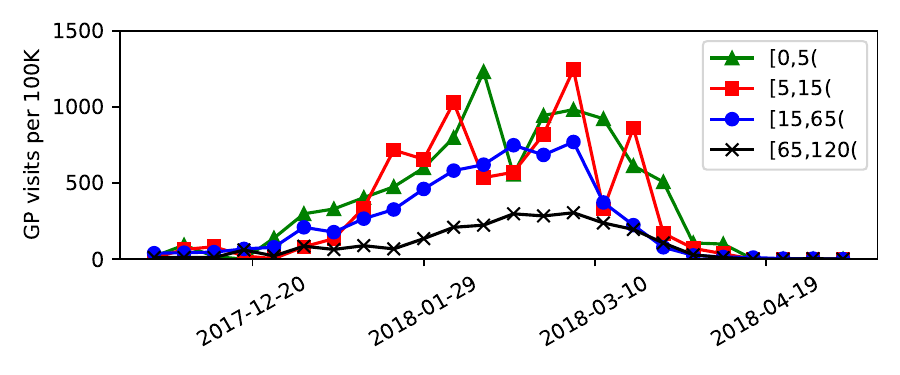}
    \caption{Weekly number of visits at GPs with influenza-Like illness per 100.000 inhabitants during the 2017-2018 season, extracted from Bossuyt et al. \citep{Bossuyt2018}.}
    \label{fig:influenza_data}
\end{figure}

\end{appendices}

\pagebreak
\bibliography{bibliography}

\end{document}